\documentclass[aps,prd,twocolumn,groupedaddress,showpacs,showkeys]{revtex4-2} 

\usepackage[utf8]{inputenc} 

\usepackage{enumitem}

\usepackage[unicode=true, pdfusetitle, bookmarks=true, bookmarksnumbered=false, bookmarksopen=false, breaklinks=true, pdfborder={0 0 0}, backref=false, colorlinks=true, linkcolor=blue, citecolor=blue, urlcolor=blue]{hyperref}
\usepackage{physics}
\usepackage{graphicx}
\usepackage{amsmath}
\usepackage{amsfonts} 
\usepackage{tensor}
\usepackage{comment}
\usepackage{color}
\usepackage[textsize=tiny,colorinlistoftodos,shadow]{todonotes}
\usepackage[normalem]{ulem}
\definecolor{hartmutdirkcorrect}{rgb}{.7,.7,.2}
\definecolor{dirkcorrect}{rgb}{.7,.2,.2}
\definecolor{dirksuggest}{rgb}{.2,.2,.7}

\usepackage{bbm}

\begin{document}

\title{A first numerical investigation of\\a recent radiation reaction model and\\ comparison to the Landau-Lifschitz model}
\author{Christian Bild, Dirk - André Deckert, Hartmut Ruhl}
\date{\today}

\begin{abstract}
    In \cite{bild2019radiation} we presented an explicit and non-perturbative derivation of the classical radiation reaction force for a cut-off modelled by a special choice of tubes of finite radius around the charge trajectories. In this paper, we provide a further, simpler and so-called reduced radiation reaction model together with a systematic numerical comparison between both the respective radiation reaction forces and the Landau-Lifschitz force as a reference. We explicitly construct the numerical flow for the new forces and present the numerical integrator used in the simulations, a Gauss-Legendre method adapted for delay equations. For the comparison, we consider the cases of a constant electric field, a constant magnetic field, and a plane wave. In all these cases, the deviations between the three force laws are shown to be small. This excellent agreement is an argument for plausibility of both new equations but also means that an experimental differentiation remains hard. Furthermore, we discuss the effect of the tube radius on the trajectories, which turns out to be small in the regarded regimes. We conclude with a comparison of the numerical cost of the corresponding integrators and find that the integrator of the reduced radiation reaction to be numerically most and the integrator of Landau-Lifschitz least efficient. 
\end{abstract}

\maketitle

\tableofcontents

\section{Introduction}

In our previous work \cite{bild2019radiation}, and as many before us, we have
reviewed Dirac's original derivation of the self-reaction force, i.e., the
Lorentz-Abraham-Dirac (LAD) equation for a single charged particle
interacting with its own retarded Liénard-Wiechert and an additional external electromagnetic field. To avoid the well-known divergences in case of
point charges, Dirac himself \cite{dirac1938classical} imposed a rigid extended charge model, parameterized by
$\epsilon>0$, which can be shrunk to a point-like charge by means of a limiting
procedure $\epsilon\to 0$. The energy-momentum conservation principle
between the kinetic degrees of freedom of the charge and the electromagnetic
field leads to
\begin{align}
    &p^{\alpha}(\tau_2)-p^{\alpha}(\tau_1)
    \label{eq_delta_momenta}
    \\
    &=q\int_{\tau_1}^{\tau_2} d\tau \,
    \left(
        {{F_\epsilon}^{\alpha}}_\beta(z(\tau))+{{F_\text{ext}}^{\alpha}}_\beta(z(\tau))
    \right) u^\beta(\tau) \, ,
    \nonumber
\end{align}
where $p^\alpha(\tau)$ denotes the four-momentum of the charge at world-line
parameter $\tau$, $q$ the total charge, and $F^{\alpha\beta}_\epsilon$,
$F^{\alpha\beta}_\text{ext}$ the field tensors of the retarded Liénard-Wiechert
field generated by the charge trajectory $\tau\mapsto z^\alpha(\tau)$ and an
external one, respectively, while $u^\beta(\tau)$ denotes the four-velocity. Based on (\ref{eq_delta_momenta}), Dirac proposed a corresponding force equation for the
charge. The latter encompasses the so-called self-force exerted on the charge
by means of its own Liénard-Wiechert field as well as the force due to the external
field. The resulting expression of the corresponding self-force is, however,
rather implicit and by virtue of Stoke's theorem involves a surface integral
over the tube $V(\tau_1,\tau_2)$ in space-time $(\mathbb{R}^4,
\eta^{\alpha\beta})$, using the signature $\eta=\operatorname{diagonal}(+1,
-1,-1,-1)$, spanned by the geometric extension of the charge along the curve segment described by
$z^\alpha(\tau)$ for
$\tau\in[\tau_1,\tau_2]$, i.e., 
\begin{align}
    \eqref{eq_delta_momenta} 
    = 
    -\int_{ \partial V(\tau_1,\tau_2) } d^3x_\beta \,
    \left(T^{\alpha\beta}_\epsilon(x) + T_\text{ext}^{\alpha\beta}(x)\right)\,.
    \nonumber
\end{align}
Here, $T^{\alpha\beta}_\epsilon$ and $T_\text{ext}^{\alpha\beta}$ denote the
respective energy-momentum tensors, and $d^3x_\beta$ the corresponding three-dimensional 3d surface measure. In order to arrive at a more tangible expression, Dirac
formally carried out a Taylor expansion in $\epsilon$ and, after subtraction
of an infinite inertial mass term, arrived at the well-known LAD equation \cite{dirac1938classical,spohn2004dynamics}:
\begin{align*}
    &\partial_\tau p^\alpha_\epsilon (\tau)
    -{{F_\text{ext}}^\alpha}_\beta u^\beta(\tau)\\
    &=
    \frac23 q^2 \left(
        \frac{da^\alpha(\tau)}{d\tau} - u^\alpha(\tau)\frac{da_\beta(\tau)}{d\tau}u^\beta(\tau) 
    \right)\,.
\end{align*}
Dirac's derivation 
is formal for two reasons: First, in the limit $\epsilon\to 0$, the field
evaluated on the charge trajectory $F^{\alpha\beta}_{\epsilon}(z(\tau))$ is
ill-defined, and thus, Stoke's theorem cannot be applied. Second, the Taylor series in $\epsilon$ is not controlled during the time integration but only at fixed instances of time. Our motivation in
\cite{bild2019radiation} to review Dirac's derivation was to avoid this Taylor
expansion entirely and, already for $\epsilon>0$, arrive at a more tangible
expression. There, we have shown in \cite{bild2019radiation} that, at least for one special choice of geometry of the
extended charge, the generated dynamics comprising the direct coupling between the Lorentz and Maxwell equations can be given in form of an explicit delay-differential equation, henceforth referred to as delayed-self-reaction (DSR) equation:

\begin{align}
    \begin{split}
        &\partial_\tau p^\alpha_\epsilon (\tau)
        -{{F_\text{ext}}^\alpha}_\beta u^\beta(\tau)\\
        = & -\frac{q^2}{6 \left[ \left( z^\gamma-z^\gamma
        (\tau-\epsilon) \right) \, u_\gamma \right]^2}\\
        & \qquad \left\{ \left[ u^\alpha-4u^\alpha(\tau-\epsilon) \, u^\beta
        u_\beta(\tau-\epsilon) \right] \right. \\
        & \qquad \left. \times \left[ 1 - u^\delta u_\delta(\tau-\epsilon)+\left( z^\rho-z^\rho
        (\tau-\epsilon) \right) \, a_\rho \right] \right\} \\
        & \qquad -
        \frac{q^2}{6 \left( z^\gamma-z^\gamma(\tau-\epsilon) \right) \,
        u_\gamma}  \\
        & \qquad \left\{ 4u^\alpha
        (\tau-\epsilon) \left[ a^\tau(\tau-\epsilon) \, u_\tau+u^\zeta
        (\tau-\epsilon) \, a_\zeta \right] \right.  \\
        & \qquad \left. +4a^\alpha(\tau-\epsilon) \, u^\vartheta(\tau-\epsilon) \,
        u_\vartheta-a^\alpha \right\} \\
        & \qquad  +\frac{2q^2}{3} \, a^\varphi(\tau-\epsilon) \, a_\varphi(\tau-\epsilon) \,
        u^\alpha(\tau-\epsilon)  =: L^\alpha_\epsilon(\tau)\,. 
    \end{split}
    \label{endgl}
\end{align}

In the following, we give a short review of these novel equation of motion, introduce a further in an ad-hoc fashion simplified version, referred to as the reduced DSR model, and conduct a numerical study of these dynamical systems while comparing to the well-known Landau-Lifschitz (LL) model in settings relevant to phenomena of radiation damping. As show in \cite{spohn2004dynamics}, the LL dynamics result from 
singular perturbation theory of the consistently coupled Maxwell-Lorentz system
in sufficiently regular external fields. In certain regimes, it is to be regarded
as a rigorous approximation of the self-force equation even in the point-charge
limit, and therefore, as a meaningful reference.

\subsection{Delayed-self force equations}\label{sec_dsr_force}

\begin{figure}
  \includegraphics[width=\linewidth*2/3]{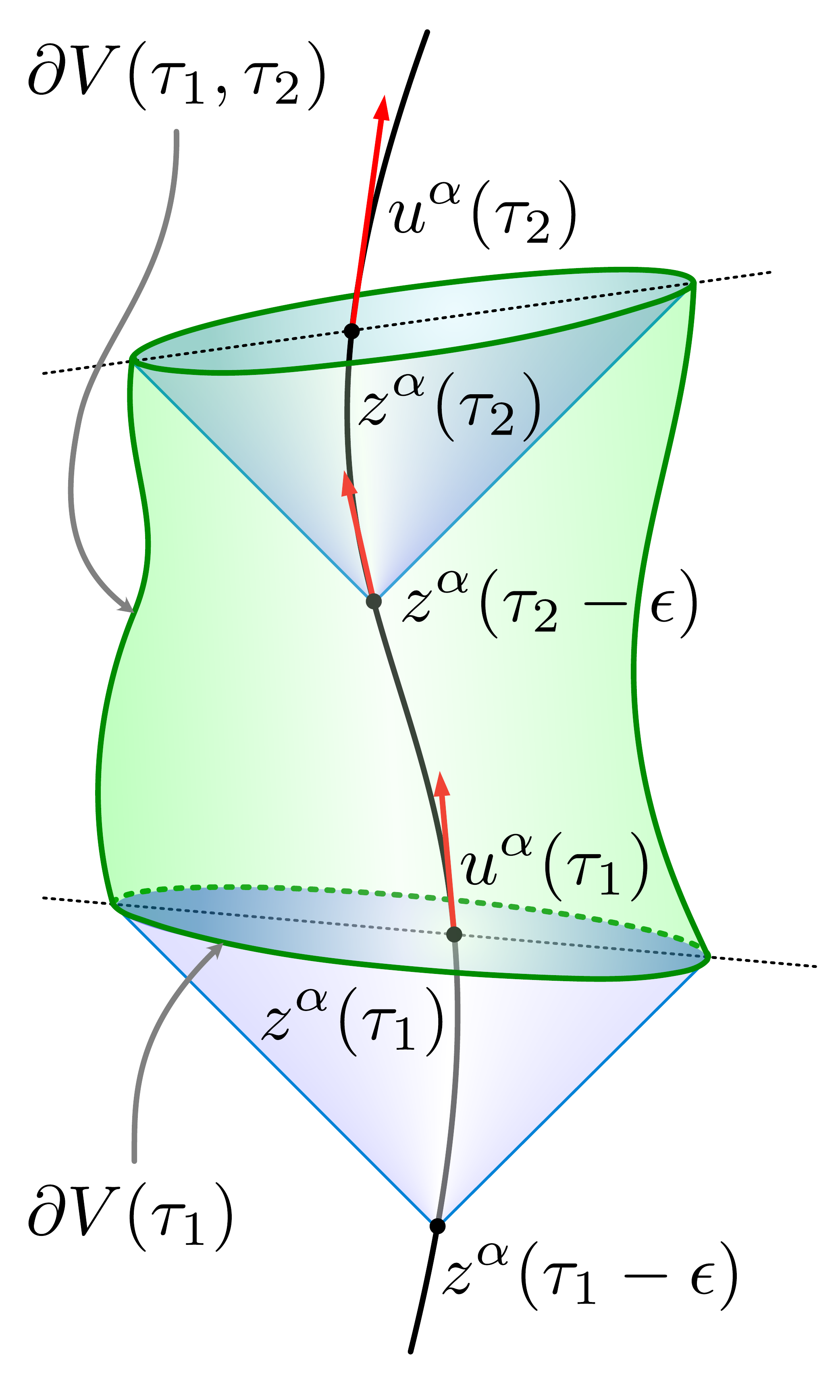}
    \caption{\label{fig_tube}The figure illustrates the tube around the charge trajectory
    $t\mapsto z^\alpha(\tau)$ which makes up the charge model employed to infer the DSR equation~\eqref{endgl}.}
\end{figure}

For the sake of self-containedness, we shall give a brief review the DSR dynamics. 
Figure~\ref{fig_tube} illustrates the special geometry
of the extended charge on the tube employed to derive the DSR equation for the segment $z^\alpha(\tau)$ of the charge trajectory for
$\tau\in[\tau_1,\tau_2]$.
For given parameter $\epsilon>0$, the geometry of the extended charge at
$\tau$ is given by the intersection $\partial V(\tau)$ of the forward
light-cone located at $z^\alpha(\tau-\epsilon)$ with the equal-time
hypersurface through $z^\alpha(\tau)$, which is four-perpendicular to the velocity
$u^\alpha(\tau)$.
The charge model is then defined by requiring that the support of the generated
charge current density $j^\alpha_\epsilon(x)$ equals the surface of the
3d set $\partial V(\tau)$ and the field fulfills
\begin{align*}
    F^{\alpha\beta}_\epsilon(x)
    =
    \begin{pmatrix}
        0 \text{ for }x\in V(\tau)\\
        F_\text{LW}^{\alpha\beta}[z](x) \text{ for }x\in \mathbb
        R^4\setminus V(\tau)\\
    \end{pmatrix},
\end{align*}
where $F_\text{LW}^{\alpha\beta}[z]$ is the retarded Linérd-Wiechert field of a
point charge situated on the world-line
$\tau\mapsto z^\alpha(\tau)$. We emphasize that
the charge-current density is not necessarily homogeneously distributed over the
surface of $\partial V(\tau)$ but given by
$j^\alpha_\epsilon(x)=\partial_\alpha F_\epsilon^{\alpha\beta}(x)$.
Furthermore, the charge trajectory does not necessarily intersect the caps of $\partial
V(\tau)$ in their centers. Both particularities are owed to the special choice of the
tube which, however, allows to derive the algebraically explicit expression 
for the DSR equation \eqref{endgl}. Notably, and somehow expected, for fixed $\tau$,
and small $\epsilon$ the right-hand side of \eqref{endgl} contains an inertial 
mass term as well as the LAD force
\begin{eqnarray}
   L^\alpha_\epsilon(\tau)=-\frac{q^2}{2\epsilon}a^\alpha+\frac{2q^2}{3}(\dot{a}^\alpha+a^\varphi
    a_\varphi u^\alpha)+\mathcal{O}(\epsilon)\,. \label{LADeq}
\end{eqnarray}

The inertial mass term diverges as $\epsilon\to\infty$. Here, it has to be emphasized that, even for
$\epsilon>0$, one would need to renormalize the charge's inertia. Even for
smooth charge models without divergences, coupling of the 
Maxwell and Lorentz equations alone implies a change of the inertial bare mass of the
charge \cite{bauer2001, spohn2004dynamics}, which needs to be gauged to the
experimentally measured one.  

For the DSR equation we proposed the following renormalization scheme in \cite{bild2019radiation}. As the
geometry of the tube, recall Figure~\ref{fig_tube}, is not static but dynamic,
one may expect some of the contributions to the inertia $m$ to depend
on the parameter $\tau$. Therefore, we start with the ansatz 
\begin{align*}
    p^\alpha(\tau)=m(\tau) u^\alpha(\tau).
\end{align*}
Exploiting $u_\alpha(\tau)a^\alpha(\tau)=0$ for for world lines, we identify
\begin{align}
    \frac{d}{d\tau} m(\tau) = L^\alpha_\epsilon(\tau) u_\alpha(\tau).
    \label{eq_ualpha_force}
\end{align}
Setting 
\begin{align*}
    F^\alpha_\epsilon(\tau) &= L^\alpha_\epsilon(\tau) - L^\beta_\epsilon(\tau)
    u_\beta(\tau) u^\alpha(\tau),\\
    F^\alpha_{\text{ext}}(\tau) &= q{F^{\alpha}}_{\beta}(z(\tau)) u^\beta(\tau)
\end{align*}
results in relativistic four-forces, i.e., ones being four-perpendicular to the
four-velocity of the charge. Finally, we find the following system of equations
\begin{align}\label{DSR}
    \frac{d}{d\tau}
    \begin{pmatrix}
        z^\alpha(\tau) \\
        u^\alpha(\tau) \\
        m(\tau)
    \end{pmatrix}
    =
    \begin{pmatrix}
        u^\alpha(\tau) \\
       \frac{1}{ m(\tau)} \, \left( F^\alpha_\epsilon(\tau)
            +
        F^\alpha_\text{ext}(\tau) \right) \\
        {u}_\alpha(\tau) L^\alpha_\epsilon(\tau)
    \end{pmatrix} \,,
\end{align}
which we refer to as the DSR system of equations.

For $\epsilon>0$, the DSR system (\ref{DSR}) represents a neutral system of
delay differential equations, meaning that the highest derivative of a solution
component occurs with and without delay. A general solution theory, and
especially a study of the case $\epsilon\to 0$, is still pending. Nevertheless,
candidates for initial conditions are certainly twice continuously
differentiable segments of the charge trajectory
\begin{align*}
    [-\epsilon,0]&\to\mathbb R^4, \tau\mapsto z^\alpha(\tau)
\end{align*}
together with an initial inertia $m(0)$. The static divergent inertia in \eqref{LADeq} can be renormalized by setting the initial inertial mass equal to
\begin{align}
\label{renormalization}
    m(0) = m - \frac{q^2}{2\epsilon} \, ,
\end{align}
which at the same time sets the effective initial
inertia to some experimental value $m(0)$. Note however, that $m(\tau)$ may
still be both $\tau$- as well as $\epsilon$-dependent. Nevertheless, there is numerical evidence that the corresponding solutions of the DSR system depend only weakly on
$\epsilon$ which also raises the hope that a rigorous analysis of the limit $\epsilon\to
0$ can be done. While we think, this is the case, it must be noted that the
right-hand side of \eqref{renormalization} becomes negative for $\epsilon$
below the classical electron radius and, in this case, renders the corresponding dynamics unstable; as
already seen in \cite{bauer2001} for the coupled Maxwell-Lorentz system of rigid charges.

\subsection{A reduced DSR model for small $\epsilon$}
\label{sec_reduced_model}

In this section, the goal is to formulate a simplified, yet still meaningful,
version of the system \eqref{DSR}, namely with the force (\ref{red-flow})
below, from which we can expect that the corresponding solutions sets are close in certain
regimes and $\epsilon$-ranges. The main advantage of (\ref{red-flow}) is the much
simpler structure, which reduces the numerical cost
significantly.

This simplified force can be motivated by regarding only the following, in the limit $\epsilon\to 0$ for fixed $\tau$
leading, terms:
\begin{align}\label{leadingterms}
    L_\epsilon^\alpha(\tau)\approx&-\frac{q^2}{2}a^\alpha+\frac{2q^2}{3\epsilon}(a^\alpha(\tau)-a^\alpha(\tau-\epsilon))\nonumber\\
    &+\frac{2q^2}{3} \, a^\gamma \, a_\gamma \,u^\alpha \, .
\end{align}
Now we permit us to drop all high-order terms and
even remove the part of the force in $u^\alpha$-direction in order to avoid a
$\tau$-dependend inertia coming from \eqref{eq_ualpha_force} which results in a constant
$m(\tau)=m - \frac{q^2}{2\epsilon}$. 
The lower two components of \eqref{DSR} can now be combined into the equation
\begin{align}\label{red-flow}
    \begin{split}
        m a^\alpha= &F^{\alpha}_\text{ext} +
        \frac{2q^2}{3\epsilon} a^\gamma(\tau-\epsilon) u_\gamma u^\alpha
        \\ 
        &+\frac{2q^2}{3} \frac{a^\alpha(\tau)-a^\alpha(\tau-\epsilon)}{\epsilon}\,.
    \end{split}
\end{align}
We will refer to the system of equations comprised by the first row of
\eqref{DSR} and equation \eqref{red-flow} together with the above constant choice of $m(\tau)$ as the reduced DSR system. Note the additional
advantage that the right-hand side of \eqref{red-flow} is linear in the
acceleration so that the corresponding flow can be given explicitly in contrast
to the one of \eqref{leadingterms}. In \eqref{red-flow}, we will again assume
$\epsilon>\frac{2q^2}{3m}$ \eqref{num_red_flow} to avoid the discussed dynamical instability.

In view of our objectives in formulating the reduced DSR system, the numerical cost  evaluating
\eqref{red-flow} is now considerably lower as compared to the full DSR model \eqref{DSR}, i.e.,
(\ref{mateqn}) below, or even the LL model, c.f., (\ref{lleqm}) below, as for the latter, the derivative of the field
strength tensor, (\ref{fieldder}) below, is needed in the multi-particle case. A
similar equation has been suggested in \cite{faci2016time}. Equations
(\ref{red-flow}) still lead to a change of the inertia $m$ as can be seen in
the case of uniform acceleration. Say, $F^\alpha_\text{ext}$ was tuned such that it
produces a solution with 
four-velocity taking the form
$u^\alpha=(\cosh(g\tau),\sinh(g\tau),0,0)$, we find
\begin{align*}
    &\frac{2q^2}{3\epsilon} a^\gamma(\tau-\epsilon) u_\gamma u^\alpha
    -\frac{2q^2}{3} \frac{a^\alpha(\tau-\epsilon)-a^\alpha(\tau)}{\epsilon}\\
    &=
    -\frac{4q^2}{3\epsilon}\sinh(\frac{g\epsilon}{2})^2a^\alpha\,
\end{align*}
similar to the change in inertia in the case of the DSR system discussed in
\cite{bild2019radiation}. This illustrates also that $m(\tau)$ may not simply be regarded as the total inertia of the charge.

\section{Numerical study}

The following sections detail the results of a systematic numerical study of
the full DSR model \eqref{DSR}, reduced DSR model \eqref{red-flow}, and the LL
model, c.f., \eqref{lleqm} below. First, in Section \ref{sec_flow} we bring
\eqref{DSR} into a form suitable for numerical integration. In Section
\ref{sec_numerical_method} we introduce the employed numerical integrator. In Section
\ref{sec_remark_uniqueness} we discuss the issue of uniqueness of solutions of the delayed system of the DSR model. Afterwards we turn to the numerical comparison. As there are very few explicit solutions, some of
which have been already discussed in \cite{bild2019radiation}, we base our
numerical investigation on the comparison of the dynamics of (\ref{DSR}) and
(\ref{red-flow}) with the LL dynamics as a reference, employing the physical units as described in Section \ref{sec_units}. The first main result is the excellent agreement between the full DSR, reduced DSR, and LL
solutions in given scenarios relevant to the radiation damping phenomena, which
we present in Section \ref{sec_comp_LL}. As a second line of investigation, we
study the $\epsilon$-dependence of both the full and reduced DSR models after renormalization. In Section \ref{sec_epsilon_dep}, as a second main result, only
a weak $\epsilon$-dependence is obtained for both the full and reduced DSR models, thus, substantiating
the discussed hope in the pending rigorous study of the limit $\epsilon\to 0$. Finally, we conclude with a summary and outlook in Section \ref{sec_outlook}, provide a sanity tests of our numerical method for all three models in  Section
\ref{sec_tests}, and we comment in detail on the numerical efficiency in Section \ref{sec_efficiency}.

\subsection{The differential flows of the full DSR and reduced DSR equations}
\label{sec_flow}

For the numerical analysis of a system of differential equations, it is helpful
to bring it into the standard form $\dot{y}(\tau)=f(y(\tau),\tau)$,
denoting the tuple of all solution components by $y(\tau)$.
We will consider delayed terms, such as $a^\alpha(\tau-\epsilon)$, which appear in the DSR systems not as extra variables but as given quantities at $\tau$ which are to be
obtained from the history of the solution. This renders $f$ in turn $y$-history-dependent. Nevertheless, as we shall demonstrate, this view can be maintained during the numerical integration as long as the integration steps are chosen smaller than $\epsilon$. The resulting function
$f(y(\tau),\tau)$ is then a vector field whose integral curves are the solution
of the differential equation. Rearranging the differential equation
into this standard form is not always possible. In our case, luckily, it can be
done since the highest derivatives $a^\alpha(\tau)$ only appear in a linear
fashion. From \eqref{DSR} we infer
\begin{eqnarray}\label{endgl}
    &&m a^\alpha+\dot{m}u^\alpha=q F^{\alpha\beta}_{ext}  u_\beta \nonumber \\
    &&-\frac{q^2}{6 \left[ \left( z^\gamma-z^\gamma
    (\tau-\epsilon) \right) \, u_\gamma \right]^2}\nonumber\\
    &&\left\{ \left[ u^\alpha-4u^\alpha(\tau-\epsilon) \, u^\beta
    u_\beta(\tau-\epsilon) \right] \right.\nonumber \\
    &&\left. \times \left[ 1 - u^\gamma u_\gamma(\tau-\epsilon)+\left( z^\gamma-z^\gamma
    (\tau-\epsilon) \right) \, a_\gamma \right] \right\} \nonumber\\
    &&-
    \frac{q^2}{6 \left( z^\gamma-z^\gamma(\tau-\epsilon) \right) \,
    u_\gamma} \nonumber \\
    &&\left\{ 4u^\alpha
    (\tau-\epsilon) \left[ a^\gamma(\tau-\epsilon) \, u_\gamma+u^\gamma
    (\tau-\epsilon) \, a_\gamma \right] \right. \nonumber \\
    &&\left. +4a^\alpha(\tau-\epsilon) \, u^\gamma(\tau-\epsilon) \,
    u_\gamma-a^\alpha \right\} \nonumber\\
    && +\frac{2q^2}{3} \, a^\gamma(\tau-\epsilon) \, a_\gamma(\tau-\epsilon) \,
    u^\alpha(\tau-\epsilon) \, ,
\end{eqnarray} 
where we suppressed the $\tau$ dependence of the non-delayed terms in our
notation. To obtain the standard form, the acceleration must be factored out on
the right-hand side of (\ref{endgl}). In the following, we split it into
vectors independent of the acceleration and products of matrices including the
acceleration. With $v^\alpha_1$ defined as 
\begin{eqnarray}
&&v^\alpha_1=-\frac{q^2}{6 \left[ \left( z^\gamma-z^\gamma
	(\tau-\epsilon) \right) \, u_\gamma \right]^2}\nonumber\\
&&\left\{ \left[ u^\alpha-4u^\alpha(\tau-\epsilon) \, u^\beta
u_\beta(\tau-\epsilon) \right] \right.\nonumber \\
&&\left. \times \left[ 1 - u^\gamma u_\gamma(\tau-\epsilon) \right] \right\} \,,\nonumber\\
\end{eqnarray}
and $k_1^{\alpha\gamma}$ defined as
\begin{eqnarray}
&&k_1^{\alpha\gamma}=-\frac{q^2}{6 \left[ \left( z^\gamma-z^\gamma
	(\tau-\epsilon) \right) \, u_\gamma \right]^2}\nonumber\\
&&\left\{ \left[ u^\alpha-4u^\alpha(\tau-\epsilon) \, u^\beta
u_\beta(\tau-\epsilon) \right] \right.\nonumber \\
&&\left. \times  \left( z^\gamma-z^\gamma
(\tau-\epsilon) \right) \right\} \,,\nonumber\\
\end{eqnarray}
the first fraction in (\ref{endgl}) is given by
\begin{eqnarray}
&&-\frac{q^2}{6 \left[ \left( z^\gamma-z^\gamma
	(\tau-\epsilon) \right) \, u_\gamma \right]^2}\nonumber\\
&&\left\{ \left[ u^\alpha-4u^\alpha(\tau-\epsilon) \, u^\beta
u_\beta(\tau-\epsilon) \right] \right.\nonumber \\
&&\left. \times \left[ 1 - u^\gamma u_\gamma(\tau-\epsilon)+\left( z^\gamma-z^\gamma
(\tau-\epsilon) \right) \, a_\gamma \right] \right\} \nonumber\\
&&=v_1^\alpha+k_1^{\alpha\gamma}a_\gamma\,.
\end{eqnarray} 
With $v^\alpha_2$ defined as
\begin{eqnarray}
v^\alpha_2&=& \frac{q^2}{6 \left( z^\gamma-z^\gamma(\tau-\epsilon) \right) \,
	u_\gamma} \nonumber \\
&&\left\{ 4u^\alpha
(\tau-\epsilon) a^\gamma(\tau-\epsilon) \, u_\gamma \right.\nonumber\\
&&\left.+4a^\alpha(\tau-\epsilon) \, u^\gamma(\tau-\epsilon) \,
u_\gamma \right\}\,,
\end{eqnarray} 
and $k_2^{\alpha\gamma}$ defined as
\begin{eqnarray}
&&k_2^{\alpha\gamma}=\frac{q^2}{6 \left( z^\gamma-z^\gamma(\tau-\epsilon) \right) \,
	u_\gamma} \nonumber \\
&&\left\{ 4u^\alpha(\tau-\epsilon) u^\gamma(\tau-\epsilon) -\eta^{\alpha\gamma} \right\}\,,
\end{eqnarray}
the second fraction in (\ref{endgl}) is given by
\begin{eqnarray}
&&\frac{q^2}{6 \left( z^\gamma-z^\gamma(\tau-\epsilon) \right) \,
	u_\gamma} \nonumber \\
&&\left\{ 4u^\alpha
(\tau-\epsilon) \left[ a^\gamma(\tau-\epsilon) \, u_\gamma+u^\gamma
(\tau-\epsilon) \, a_\gamma \right] \right. \nonumber \\
&&\left. +4a^\alpha(\tau-\epsilon) \, u^\gamma(\tau-\epsilon) \,
u_\gamma-a^\alpha \right\} \nonumber\\
&&=v_2^\alpha+k_2^{\alpha\gamma}a_\gamma\,.
\end{eqnarray}
The remaining part of (\ref{endgl}), which could be coined the delayed Lamor
formula, is independent of the acceleration and we define it as $v^\alpha_3$.
With these abbreviations, (\ref{endgl}) reduces to 
\begin{equation}\label{abre}
(m \eta^{\alpha\gamma}-k_1^{\alpha\gamma}-k_2^{\alpha\gamma})a_\gamma+\dot{m}u^\alpha=v_1^\alpha+v_2^\alpha+v_3^\alpha\,.
\end{equation}
Since $a^\alpha=\dot{u}^\alpha$, (\ref{endgl}) can be understood as four
equations containing nine variables $z^\alpha,u^\alpha$ and $m$. In
addition to the four equations in (\ref{endgl}), the identity
$\dot{z}^\alpha=u^\alpha$, and the constraint $u^\alpha a_\alpha=0$ determine
the time evolution of all nine variables completely. With $$k^{\alpha\gamma}=m
\eta^{\alpha\gamma}-k_1^{\alpha\gamma}-k_2^{\alpha\gamma}$$ and
$v^\alpha=v_1^\alpha+v_2^\alpha+v_3^\alpha$, the equation (\ref{abre}) and the constraint $u^\alpha a_\alpha=0$ can be recast into the matrix equation
\begin{equation}\label{mateqn}
    A \begin{pmatrix}
        \dot{m}\\
        a^\gamma\\
    \end{pmatrix}=
    \begin{pmatrix}
        0\\
        v^\alpha\\
    \end{pmatrix}\,,
    \qquad
    A=\begin{pmatrix}
        0 & u_\gamma\\
        u^\alpha &\tensor{k}{^\alpha_\gamma}
    \end{pmatrix}\,,
\end{equation}
One obtains the differential flow by inverting matrix in \eqref{mateqn}. The existence of the inverse is discussed in chapter \ref{sec_remark_uniqueness}. 
The resulting system of nine variables is then given by
\begin{eqnarray}
\frac{d}{d\tau} z^\alpha&=&u^\alpha\label{trivial}\,,\nonumber\\
\frac{d}{d\tau}
\begin{pmatrix}
m\\
u^\gamma\\
\end{pmatrix}&=&
A^{-1}
\begin{pmatrix}
0\\
v^\alpha\\
\end{pmatrix} \,.
    \label{eq_matrix_A_inverse}
\end{eqnarray}
This explicit expression of the differential flow simplifies the
numerical analysis considerably.

Regarding the reduced DSR model, the corresponding differential flow is obtained from 
\eqref{red-flow} by solving for the highest-order derivative that depends on $\tau$:
\begin{align}\label{num_red_flow}
    \frac{d}{d\tau}u^\alpha =&\frac{1}{m-\frac{2q^2}{3\epsilon}}\Big(qF^{\alpha\beta}_{ext}u_\beta \\\nonumber
    &+\frac{2q^2}{3\epsilon}\left(a^\gamma(\tau-\epsilon) u_\gamma u^\alpha-a^\alpha(\tau-\epsilon)\right)\Big)\,.
\end{align}
We alert the reader to the denominator. In order to avoid the discussed dynamical instability we restrict our study to $\epsilon > \frac{2q^2}{3m}$.

\subsection{The numerical integrator}
\label{sec_numerical_method}
Various methods to numerically integrate ordinary differential equations are
known; see, e.g., \cite{hairer2006geometric,iserles2009first}. To be
self-contained, we give a brief review of the employed integration method
suitable to also treat the delayed terms. In the abstract
setting of differential equations without delay, i.e., for the class
\begin{equation}\label{basic} 
\dot{y}(t)=f(y(t),t),
\end{equation} 
using time parameter $t$ instead of $\tau$,
with
initial value $y(T)=y_0$ at some initial time $T$, it is the objective of the
numerical integration to calculate a sequence of the solution values $y(T+nh)$,
$n=0,1,2, \dots$, for a given integration step size $h>0$. As mentioned earlier, we want to regard the delayed terms that also contribute to $f(y(t),t)$, such as $y(t-\epsilon)$ and its derivatives, not as extra variables but fixed quantities that are obtained from the history of the solution. For this purpose, we need to choose $h<\epsilon$ in order to obtain a reliable interpolation on the scale $\epsilon$ and keep the delayed
values accessible to the numerical integrator -- the suppression of the dependence of $f$ on the delayed terms in our notation in \eqref{basic} is therefore not misleading in such a setting, though one should keep in mind that $f$ always depends on the respectively last $\epsilon$-segment of the solution history. In principle, one could simply use a step size
which is given by the delay $\epsilon$
divided by a natural number. But in this case, a dynamical adjustment of the
step size, which can be very beneficial for the integrator as discussed below, would be impossible. A class of methods, which allow the calculation
of interpolation values without further assumptions are the so-called collocation
methods. The central idea behind these methods is the approximation of the
trajectory with a conveniently chosen polynomial. For this purpose, it is helpful to parameterize 
$t=T+hc:=t_c$ by means of a unitless quantity $c$ in $[0,1]$ for a given
integration step $h>0$. Using this parameterization, equation~(\ref{basic}) can be recast into the
following integral form 
\begin{equation}\label{intequ}
    y(t_{c})=y_0+h\int_0^{c}f(y(t_r),t_r) dr\,,
\end{equation}
A direct integration of (\ref{intequ}) is obviously not possible since the integrand
depends on the unknown $y(t)$. Instead, the equation must be read as a self-map
on a space solution candidates. For our numerical approach of solving for
$y(t)$, it is convenient to approximate the integrand $f(y(t),t)$ by a
polynomial $p(t)$ of a certain degree $n-1$. In order to sample such a
polynomial, we define the sample values $k_j$ at collocation points $t_j$ 
\begin{align*}
    k_j=f(y(t_j),t_j),
    \qquad
    t_j\equiv t_{c_j}=T+c_j h
\end{align*}
for $j=1,2,...,n$, where the $c_j$ specify the distribution of the $n$
collocation points in units of $h$. Optimal values of the latter will be
discussed below. With the help of the corresponding Lagrange polynomials
\begin{equation}
    l_j(t)=\prod_{\iota\neq j}\frac{\frac{t-T}{h}-c_\iota}{c_j-c_\iota}
\end{equation}
for $j=1,2,\dots,n$, which fulfill $l_j(t_i)=\delta^i_j$, the interpolation
polynomial $p(t)$ can be given in terms of
\begin{equation}\label{interpol}
    p(t)=\sum^{n}_{j=1} k_j l_j(t)\,.
\end{equation}
By construction, we have $p(t_i)=k_i$ and we may now use $p(t)$ as an
approximation of $f(y(t),t)$. At the collocation points, the polynomial
interpolation and the equation~\eqref{intequ} implies 
\begin{align}
    \xi_i:= &y(t_i) = y_0 + h \int_0^{c_i}f(y(t_r), t_r) dr\nonumber\\
    &\approx y_0 + h \int_0^{c_i}p(t_r) dr
    =y_0 + h\sum_{j=1}^n a_{ij} k_j,
    \label{xi}
\end{align}
where the $a_{ij}$ coefficients are given by
\begin{equation}
    a_{ij}=\int^{c_i}_0 l_j(t_r) dr\,.
\end{equation}
In this way, we obtain a discreet version of the self-map \eqref{intequ} in
terms of the variables $(\xi_i,k_i)_{i=1,\dots n}$ that
reads
\begin{align}
    \label{eq_discreet_self_map}
    \begin{pmatrix}
        \xi_i\\
        k_i
    \end{pmatrix}_{i=1,\dots,n}
    =
    \begin{pmatrix}
        y_0 + h\sum_{j=1}^n a_{ij} k_j\\
        f(\xi_i, t_i)
    \end{pmatrix}_{i=1,\dots,n}\,,
\end{align}
which may be attempted to be solved for a fixed-point iteratively. As stressed
before, for this to work in our setting incorporating the delay $\epsilon$, the
step size $h$ has to be sufficiently smaller than $\epsilon$ as $f(\cdot,t)$ can
only be computed on the basis of these delayed terms.

In case of convergence of this iteration, the corresponding speed will depend
on the step size $h$, which is larger
for smaller $h$. A good initial choice can
reduce the numerical cost of the fixed-point procedure. E.g., we have used 
$$z(\tau)+hu(\tau), \quad u(\tau)+ha(\tau), \quad a(\tau), \quad \text{and}
\quad m(\tau)$$ 
as starting point for the respectively next fixed-point approximation. Finally, the next point
$y(t+h)$ can again be obtained through (\ref{intequ}), i.e., by
\begin{align}\label{next_step}
y(t+h)
=y_0+h\sum^n_{i=1}k_i\int_0^{1}l_i(t)\, dt
= y_0+h\sum^n_{i=1} k_i b_i\,,
\end{align}
for coefficients $b_i$ given by $ b_i=\int^1_0l_i(t)\,dt.
$
The quality of the approximation of the numerical procedure depends on the
choice of the $c_i$. In general, it can be proven that the order of the collocation method
$n$ is at least equal to the order of approximation of the integral in the step
size $h$, i.e.,
\begin{equation}
	\int_{T}^{T+h}f(y(t),t)\,dt=\int_{T}^{T+h}p(t)\,dt+\mathcal{O}(h^{n}).
\end{equation}
The numerical approximation of an integral by means of this polynomial
approximation is called quadrature and it is well-known that for a convenient
choice of collocation points $t_i$, $i=1,2,\dots,n$, one can obtain an
approximation error of order $2n$ by the following procedure.
Let us consider, the shifted Legendre polynomials given by
\begin{equation}
	p_k(t)=\frac{1}{k!}\frac{d^k}{dt^k}t^k(t-1)^k\,,
\end{equation}
which fulfill the orthogonality condition
\begin{equation}
    \int_{0}^{1}p_i(t)p_j(t)dt=\frac{\delta^i_j}{2i+1}
\end{equation}
and form a basis of the space of polynomials on $[0,1]$. Any polynomial $p(t)$
of order smaller or equal $2n-1$ can be written in the form
\begin{equation}
    \label{eq_gl_type}
	p(t)=q(t)\,p_n(t)+r(t)\,,
\end{equation}
for a polynomial $q(t)$ of order smaller or equal $n-1$ of the form
\begin{equation}
	q(t)=\sum_{i=0}^{n-1}q_ip_i(t)\,,
\end{equation}
and with corresponding coeficients $q_i$ and a remainder polynomial $r(t)$, which
is consequently also of order smaller or equal $n-1$.
The advantage such a decomposition of $p(t)$ is that, thanks to the
orthogonality of the Legendre polynomials, an integration reduces to
\begin{eqnarray}
	\int_{0}^{1}p(t)\,dt=\int_{0}^{1}\left(\sum_{i=0}^{n-1}q_ip_i(t)p_n(t)+r(t)\right)\,dt\nonumber\\=\int_{0}^{1}r(t)\,dt.
\end{eqnarray}
It remains to determine $r(t)$ without having to find $p(t)$ first. For this,
one exploits the fact that $p_n(t)$ has exactly $n$ roots in the interval
$[0,1]$, which is a well-known consequence of the orthogonality conditions and the
intermediate value theorem. Choosing the collocation points
$t_1, t_2,\dots, t_n$ to lie exactly at the $n$ roots of the Legendre polynomial $p_n(t)$ implies
$$k_i=p(t_i)=r(t_i)\,,$$ 
which determines $r(t)$ uniquely.
The integration strategies resulting from the special form of the $a_{ij}$, $b_i$, and the ansatz \eqref{eq_gl_type} for $p(t)$ with collocation points taken as the roots of the Legendre polynomial $p_n(t)$ are
called Gauss-Legendre methods \cite{hairer2006geometric,iserles2009first}, which are a sub-category of the Runge-Kutta methods. Note
that $p(t)$ does not need to be specified explicitly.
Instead, it suffices to specify the coefficients $a_{ij},b_i$ and
$c_i$, which is usually done in so-called Butcher tableaus:
\def\arraystretch{1.2}
\begin{equation}
\begin{array}{c|c}
c_i&a_{ij}\\
\hline
&b_i
\end{array}\,.
\end{equation}
By virtue of \eqref{eq_discreet_self_map}, these coefficients contain all
information needed to implement the integration procedure. For instance, for
$n=1$ and $n=2$ the Butcher tableaus are given by
\begin{equation}
    \begin{array}{c|c}
        \frac{1}{2}&\frac{1}{2}\\
        \hline
        &1
    \end{array}
    \quad\text{and}\quad
    \begin{array}{c|c c}
        \frac{1}{2}-\frac{\sqrt{3}}{6}&\frac{1}{4}&\frac{1}{4}-\frac{\sqrt{3}}{6}\\
        \frac{1}{2}+\frac{\sqrt{3}}{6}&\frac{1}{4}+\frac{\sqrt{3}}{6}&\frac{1}{4}\\
        \hline
        &\frac{1}{2}&\frac{1}{2}
    \end{array}\,,
\end{equation}
respectively, which achieve an approximation error of order $2n$ in $h$. It can be proven for all collocation methods, including the Gauss-Legendre method discussed above, that this is the optimal approximation error. The Gauss-Legendre methods are furthermore
A-stable, B-stable, symmetric, symplectic and conserve quadratic invariants
\cite{hairer2006geometric,iserles2009first}. 

\subsection{Remark on the uniqueness of DSR solutions}
\label{sec_remark_uniqueness}

Due to the delayed terms in the differential equations (\ref{DSR}), initial
values $z(0), u(0), m(0)$ do not suffice to integrate it. At least an entire
segment of the initial trajectory, say, $\tau\mapsto(z(\tau),u(\tau),a(\tau))$
for $\tau$ in $[-\epsilon,0]$, is necessary to integrate the right-hand side of
(\ref{DSR}) forward in $\tau$. 
As discussed in Section~\ref{sec_numerical_method}, one approach to find
solutions is to recast \eqref{DSR} into \eqref{eq_matrix_A_inverse} with the
help of the inverse of the matrix $A$, see \eqref{mateqn}. The advantages of
\eqref{eq_matrix_A_inverse} is that the highest non-delayed derivative $a(\tau)$ is on the left-hand
side without delay and everything on the right is either delayed or depends
on lower-order derivatives. Delay differential equations of this type can be
integrated by the method-of-steps as described in, e.g., 
\cite{driver2012ordinary}. Given a sufficently regular flow, they give rise
to solutions characterized uniquely by the initial data as described above. Let
us therefore investigate the invertiblity of $A$, which would then imply the discussed
sense of uniqueness of solutions to \eqref{DSR}.

The determinant of the matrix $A$ is a Lorentz scalar since it is
constructed with the help of Lorentz tensors only. A representation of the
determinant $\det(A)$, which makes the transformation properties transparent,
is obtained from the Cayley–Hamilton theorem, which states that every square
matrix fulfills its own characteristic equation.
The determinant of a five by five matrix A is given by
\begin{eqnarray}
	\det(A)&=&\frac{1}{120} \Big(-Tr(A)^5+10 Tr(A)^3 Tr(A^2)\nonumber\\
	&&-20 Tr(A)^2 Tr(A^3)-15 Tr(A) Tr(A^2)^2\nonumber\\
	&&+30 Tr(A) Tr(A^4)+20 Tr(A^2) Tr(A^3)\nonumber\\
	&&-24 Tr(A^5) \Big).
\end{eqnarray}
For our matrix this is equivalent to
\begin{eqnarray}\label{cal}
	&&\det(A)=\nonumber\\&&
	- \frac{1}{6} \tensor{k}{^\alpha_\alpha}\tensor{k}{^\beta_\beta}\tensor{k}{^\gamma_\gamma} + \frac{1}{120} \tensor{k}{^\alpha_\alpha}\tensor{k}{^\beta_\beta}\tensor{k}{^\gamma_\gamma}\tensor{k}{^\delta_\delta}\tensor{k}{^\mu_\mu}\nonumber\\&& - \frac{1}{12} \tensor{k}{^\alpha_\alpha}\tensor{k}{^\beta_\beta}\tensor{k}{^\gamma_\gamma}\tensor{k}{^\delta^\mu}\tensor{k}{_\mu_\delta} + \frac{1}{6} \tensor{k}{^\alpha_\alpha}\tensor{k}{^\beta_\beta}\tensor{k}{^\gamma^\delta}\tensor{k}{_\delta^\mu}\tensor{k}{_\mu_\gamma}\nonumber\\&& + \frac{1}{2} \tensor{k}{^\alpha_\alpha}\tensor{k}{^\beta_\beta}\tensor{k}{^\gamma^\delta}\tensor{u}{_\gamma}\tensor{u}{_\delta} + \frac{1}{2} \tensor{k}{^\alpha_\alpha}\tensor{k}{^\beta^\gamma}\tensor{k}{_\gamma_\beta}\nonumber\\&& + \frac{1}{8} \tensor{k}{^\alpha_\alpha}\tensor{k}{^\beta^\gamma}\tensor{k}{_\gamma_\beta}\tensor{k}{^\delta^\mu}\tensor{k}{_\mu_\delta} - \frac{1}{4} \tensor{k}{^\alpha_\alpha}\tensor{k}{^\beta^\gamma}\tensor{k}{_\gamma^\delta}\tensor{k}{_\delta^\mu}\tensor{k}{_\mu_\beta}\nonumber\\&& -  \tensor{k}{^\alpha_\alpha}\tensor{k}{^\beta^\gamma}\tensor{k}{_\gamma^\delta}\tensor{u}{_\beta}\tensor{u}{_\delta} - \frac{1}{6} \tensor{k}{^\alpha^\beta}\tensor{k}{_\beta_\alpha}\tensor{k}{^\gamma^\delta}\tensor{k}{_\delta^\mu}\tensor{k}{_\mu_\gamma}\nonumber\\&& - \frac{1}{2} \tensor{k}{^\alpha^\beta}\tensor{k}{_\beta_\alpha}\tensor{k}{^\gamma^\delta}\tensor{u}{_\gamma}\tensor{u}{_\delta} - \frac{1}{3} \tensor{k}{^\alpha^\beta}\tensor{k}{_\beta^\gamma}\tensor{k}{_\gamma_\alpha}\nonumber\\&& + \frac{1}{5} \tensor{k}{^\alpha^\beta}\tensor{k}{_\beta^\gamma}\tensor{k}{_\gamma^\delta}\tensor{k}{_\delta^\mu}\tensor{k}{_\mu_\alpha} +  \tensor{k}{^\alpha^\beta}\tensor{k}{_\beta^\gamma}\tensor{k}{_\gamma^\delta}\tensor{u}{_\alpha}\tensor{u}{_\delta}.
\end{eqnarray}
Since this expression is already quite complicated, there is not much use by stating its form after inserting the definition of $k^{\alpha\beta}$. Instead we present a Taylor series in $\epsilon$  of $\det(A)$ to discuss some of the features. With $u_\alpha\dot{a}^\alpha=-a_\alpha a^\alpha$, $k_1^{\alpha\gamma}$ is given up to first order by 
\begin{eqnarray}\label{tailor1}
k_1^{\alpha\gamma}&=&\frac{q^2}{6\epsilon^2}(u^\alpha-4(u^\alpha-\epsilon a^\alpha+1/2\epsilon^2\dot{a}^\alpha)\nonumber\\
&&\times(1-1/2\epsilon^2 a^\beta a_\beta)(\epsilon u^\gamma-1/2\epsilon^2 a^\gamma)).
\end{eqnarray}
Since $k_1^{\alpha\gamma}$ only appears as a matrix product with
$k_1^{\alpha\gamma}a_\gamma$ in \eqref{abre}, only the part not being orthogonal to $a_\gamma$ has to be taken into consideration. Specifically, the $u^\gamma$ can be dropped in (\ref{tailor1}), so the relevant part is
\begin{eqnarray}
{k_1^{\alpha\gamma}}_{rel}&=&\frac{q^2}{12}(3u^\alpha a^\gamma-4\epsilon a^\alpha a^\gamma).
\end{eqnarray}
More generally, contributions in direction of $u_\gamma$ can be dropped from
$k^{\alpha\gamma}$ since the first row of $A$ contains $u_\gamma$, and hence, it can
be subtracted from every other row with an arbitrary factor without changing
the determinant. Similarly, the relevant part of $k_2^{\alpha\gamma}$ to first order is given by
\begin{eqnarray}
{k_2^{\alpha\gamma}}_{rel}&=&\frac{q^2}{6\epsilon}(-\eta^{\alpha\gamma}-4\epsilon u^\alpha a^\gamma\nonumber\\
&&+4\epsilon^2 (a^\alpha a^\gamma+1/2u^\alpha \dot{a}^\gamma)).
\end{eqnarray}
With these expressions, the relevant part of $k^{\alpha\gamma}$ up to first order in $\epsilon$ is given by
\begin{eqnarray}\label{krel}
{k^{\alpha\gamma}}_{rel}&=&\frac{q^2}{12}((\frac{2}{\epsilon}+\frac{12m}{q^2})\eta^{\alpha\gamma}+5 u^\alpha a^\gamma+\nonumber\\
&&-4\epsilon (a^\alpha a^\gamma+u^\alpha \dot{a}^\gamma)).
\end{eqnarray}
Since \eqref{krel} contains negative powers of $\epsilon$ it is necessary to
take higher orders than considered in \eqref{krel} into account to evaluate
\eqref{cal}. We obtain
\begin{eqnarray}\label{expandeddet}
\det(A)&=&\frac{1}{216}\frac{1}{\epsilon^3}q^6   - \frac{1}{12}\frac{1}{\epsilon^2}q^4 m \nonumber\\&& - \frac{1}{144}\frac{1}{\epsilon}q^6  \tensor{a}{^\alpha}\tensor{a}{_\alpha} + \frac{1}{2}\frac{1}{\epsilon}q^2 m^2  \nonumber\\&& + \frac{1}{12}q^4 m \tensor{a}{^\alpha}\tensor{a}{_\alpha} + \frac{47}{5184}q^6  \tensor{a}{^\alpha}\tensor{\dot{a}}{_\alpha}\nonumber\\&& - m^3+\mathcal{O}(\epsilon)\,.
\end{eqnarray}
For uniform acceleration, given by
\begin{eqnarray}
u^\alpha&=&(\cosh(\tau),0,0,\sinh(\tau))\,,\nonumber\\
q&=&\frac{1}{10}\,,\nonumber\\
m&=&1-\frac{q^2}{2\epsilon}\,,
\end{eqnarray}
we obtain
\begin{eqnarray}
\det(A)=&&\\\nonumber
&&\frac{128 - 57600 \epsilon + 8640021 - 432003600 \epsilon^3}{ 432000000\epsilon^3}\\\nonumber
&&+\mathcal{O}(\epsilon)\,.
\end{eqnarray}
This determinant, to order as indicated in \eqref{expandeddet}, approaches zero only for
$\epsilon\approx0.00660133\dots$, which is of the order of the classical electron
radius $r_{e}=q^2/m_e\approx1/100$
, and for which we expect a dynamical
instability already due to the change in sign in the initial inertia
\eqref{renormalization}
. In our simulations, numerical instabilities also already
appear for values of $\epsilon$ a little bigger
than the classical electron
radius, which is why we cannot probe the solutions for arbitrary small
$\epsilon$. For bigger values of $\epsilon$, however, the matrix $A$ is invertible and the analysis above indicates that uniqueness of solution in the discussed sense can be expected.

\subsection{The units for the simulations}
\label{sec_units}

Regarding the numerical simulations, it is of advantage to make sure that all relevant
quantities are of the same order of magnitude, otherwise numerical accuracy
loss due to addition may become unnecessarily large. We employ the usual choices
$c=4\pi\epsilon_0=\mu_0/4\pi=1$. Also, the mass of the electron $m_e$
is set to equal $1$. The additional choice for the elementary charge $e=1/10$ then
determines the basic units for time and distance uniquely, similarly as in the case
of natural units. A unit of time is, hence, given by
\begin{equation}\label{time}
	\frac{100e^2\mu_0}{4\pi c m_e}\approx9.3996\times10^{-22} \,\text{s}\,,
\end{equation}
a distance of unit one corresponds to
\begin{equation}
	\frac{100e^2\mu_0}{4\pi m_e}\approx2.8179\times 10^{-13} \,\text{m}\,,
\end{equation}
a magnetic field of unit one is
\begin{equation}\label{mag}
	\frac{4\pi m_e^2c}{1000e^3\mu_0}\approx6.0488\times10^8\, \text{T}\,,
\end{equation}
and an electric field of unit one is given by
\begin{equation}
	\frac{4\pi m_e^2c^2}{1000e^3\mu_0}\approx1.8134\times10^{17}\, \text{V/m}\,.
\end{equation}
Mass renormalization is also possible by choosing different units instead of
using \eqref{renormalization}. The effective mass is not $m$ but 
\begin{equation}
    m_\text{eff}=m+\frac{q^2}{2\epsilon}\,.
\end{equation}
By keeping $m=1$, mass renormalisation can also be achieved by setting the
electron mass $m_e$ to $m_\text{eff}$. Notably, with the choice of
$m_e=m_\text{eff}$, the distance $\epsilon$ for $q=1/10$ and $m=1$ is given by
\begin{eqnarray}
\epsilon \frac{100e^2\mu_0}{\frac{4\pi m_e}{ 1+\frac{1}{200\epsilon}}}=
    \frac{(100\epsilon+\frac{1}{2})e^2\mu_0}{4\pi m_e}\,,
\end{eqnarray}
and this expression converges to a finite value in the limit
$\epsilon\rightarrow 0$, which is given by 
\begin{equation}
    \frac{1}{2}\frac{e^2\mu_0}{4\pi m_e}=\frac{1}{2}r_e\,,
\end{equation}
where $r_e$ is the classical electron radius. The $1/2$ stems from the fact
that the charge is located only on the surface of the
particle in our model. This means, with this renormalization scheme, $\epsilon$
is effectively bounded below by the classical electron radius. As emphasized earlier in the case of \eqref{renormalization}, smaller values for $\epsilon$ lead to a negative mass,
which provokes numerical instabilities.

\subsection{Main Result I: Comparison of DSR, reduced DSR, and LL dynamics}
\label{sec_comp_LL}

\begin{figure}
  \includegraphics[width=\linewidth]{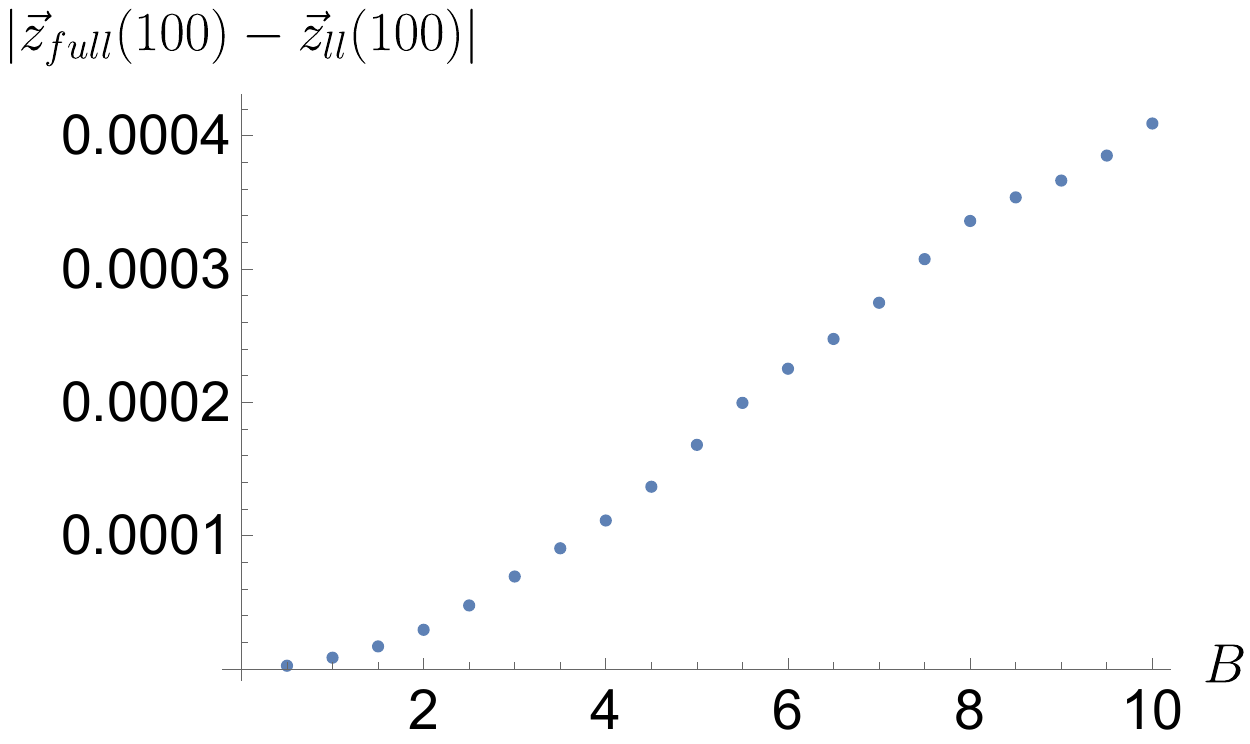}
  \caption{The figure shows the absolute value of the difference of the
    3d-positions $\vec{z}(\tau)$ for different magnetic field strengths from 0.5
    to 10 between the explicit solution to the LL equation and the numeric solution to the full DSR equation after a time of 100 with identical initial conditions.}
  \label{mag_ll_full_dif}
\end{figure}
\begin{figure}
  \includegraphics[width=\linewidth]{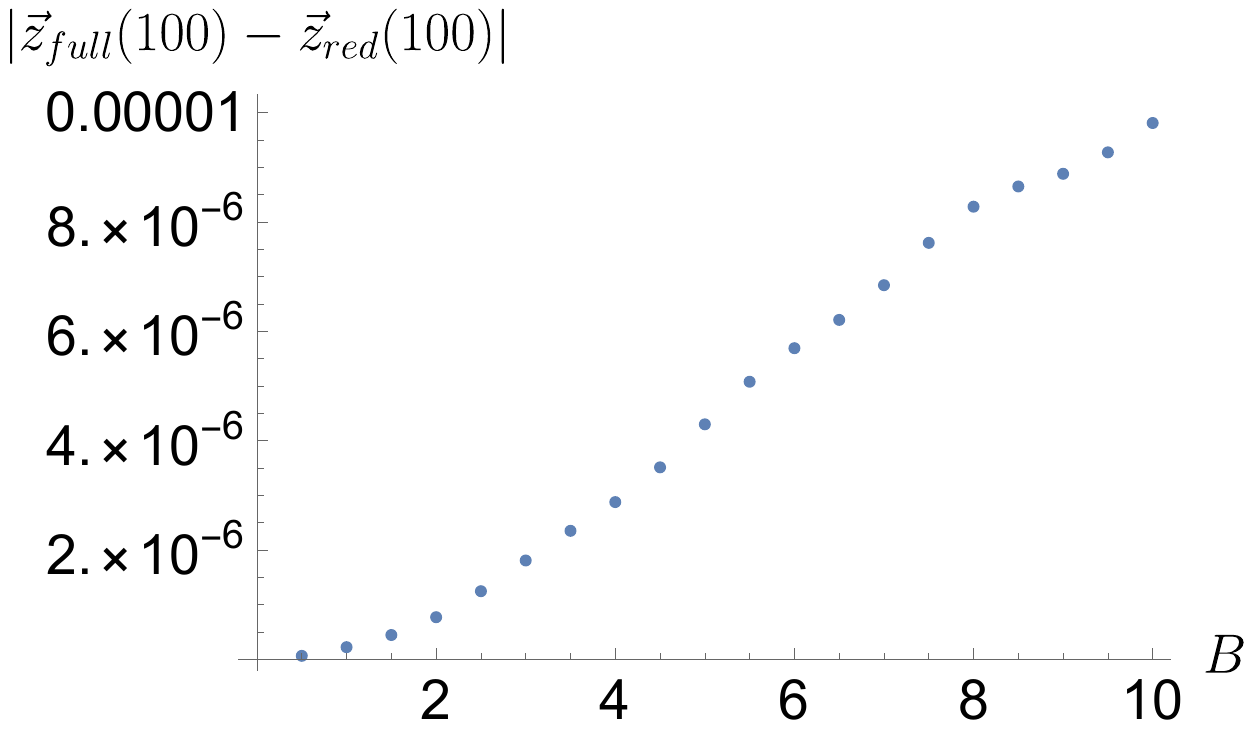}
  \caption{The figure shows the absolute value of the difference of the 3d-positions $\vec{z}(\tau)$ for different magnetic field strengths from 0.5 to 10 between the numeric solution to the full DSR equation and the numeric solution to the reduced DSR equation after a time of 100 with identical initial conditions.}
  \label{mag_full_red_dif}
\end{figure}

Our first main result is a comparison between the DSR \eqref{DSR}, reduced DSR \eqref{red-flow},  and
LL \eqref{lleqm} dynamics in the settings of a) a constant magnetic field and b) a monochromatic transverse wave.

\paragraph{Constant magnetic field:} In this setting we consider the external force $F_\text{ext}^\alpha$ to result from a constant magnetic field in $z$-direction for varying magnitudes. Luckily, for this case, there is an explicit solution to the LL equations available \eqref{ll-const-mag}. For the initial four-velocity $u^\alpha=(0,0.1,0,0)$ and for the delay $\epsilon=0.05$ are used with a duration of $T=100$, which is large enough to allow for multiple cycles of the charge, and thus, significant accumulation of radiation reaction effects. As the initial trajectory segment for both
DSR equations, the corresponding solution segment of the LL equation is used. 

Our findings are as follows. The solutions of the reduced and full DSR equations agree to the order of $10^{-5}$ as can be seen in Figure~\ref{mag_full_red_dif}, which demonstrates that the reduced DSR dynamics are an excellent approximation of the full DSR dynamics in the considered setting.
Regarding the full DSR equation, we observe that the inertia parameter $m(\tau)$ remains almost
constant over the entire duration, with the change being $\Delta m=-1.27942 \cdot 10^{-6}$.
Figure~\ref{mag_ll_full_dif} shows the absolute value of
the difference of the 3d-positions $\vec{z}(\tau)$ between the explicit solution
of the LL equation and the numeric solution to the full DSR equation.
Notably, even for a magnetic field strength of 10 units, according \eqref{mag}, the
deviation in $\vec{z}(\tau)$ between the DSR and LL dynamics for $\tau=T$ is only of the order of $10^{-4}$.
Accordingly, the difference of the reduced DSR dynamics to the LL dynamics is almost identical to the difference between the full DSR and LL dynamics as shown in Figure~\ref{mag_ll_full_dif}. 

Hence, the trajectory produced by the DSR or reduced DSR
equation cannot even be distinguished by the eye from the LL solution shown in
Figure~\ref{const_mag_ana_ll}. The extraordinary agreement between the solutions
of the LL, the DSR and the reduced DSR equations is a strong argument for the
physical plausibility of the DSR as well as reduced DSR system of equations. 
Given the major differences
not only in the structure but also in the type of these three
equations, this level of agreement is exceptional.

\begin{figure}
  \includegraphics[width=\linewidth]{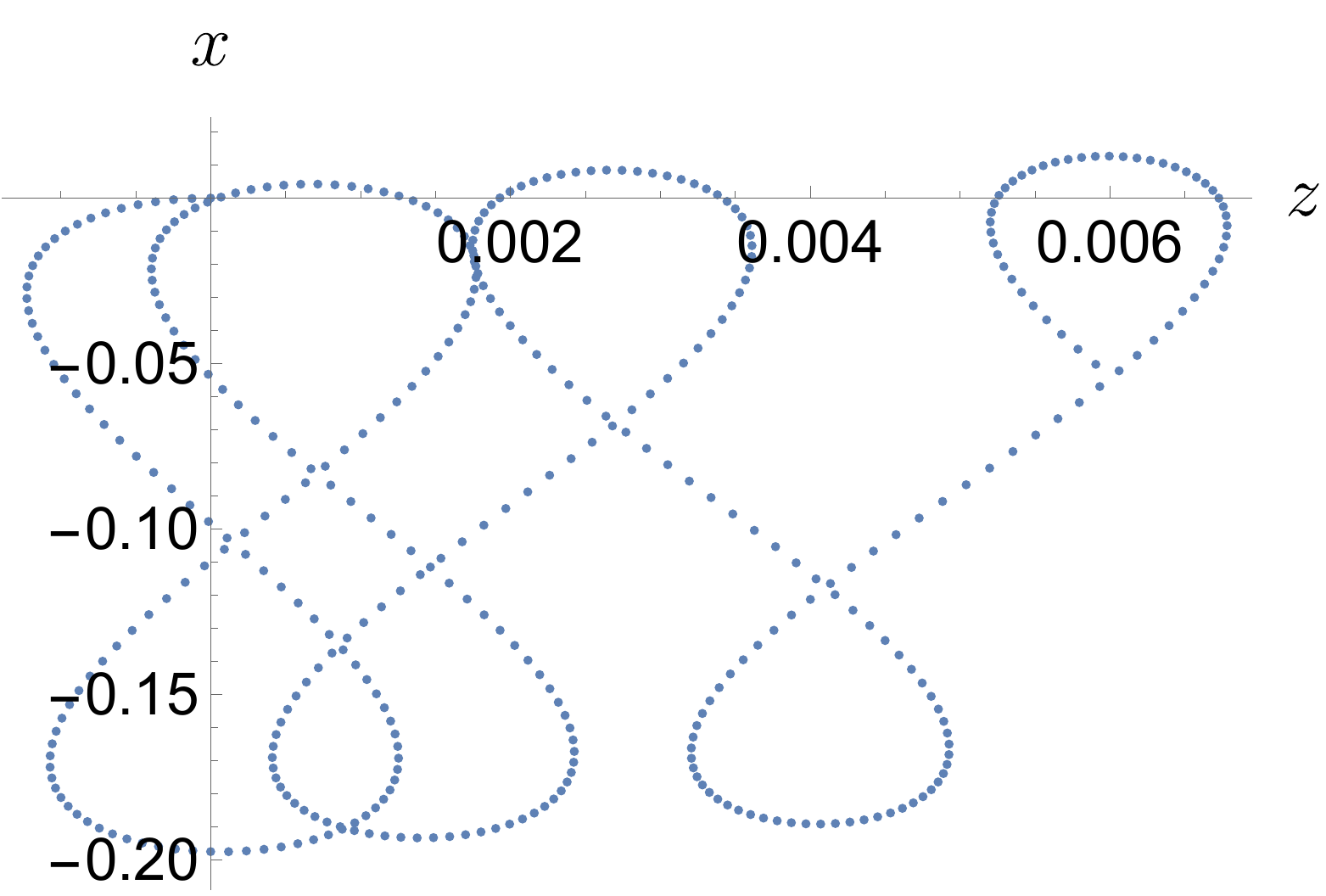}
  \caption{The figure shows the x and z coordinate of the solution to the reduced delay equation for a particle in a plane wave, which propagates in z direction.}
  \label{8red}
\end{figure}

\paragraph{Monochromatic transverse wave:} Next, a linearly polarised transverse wave given by $A^\alpha_\text{ext}=a\epsilon^\alpha \sin(n^\beta z_\beta)$ is considered, where $a$ is the field strength, $n^\alpha$ is the light-like propagation four-vector of the wave, and $\epsilon^\alpha$ is the polarisation of the wave. For our later discussion, it is useful to first consider the trajectory of a particle subject to only the Lorentz force in such a field, in particular, without radiation the reaction force, which is given by
\begin{eqnarray}
    z^\alpha(\zeta)&=&u^\alpha(0)\tau-\frac{2qa\sin^2(\zeta/2)}{m n^\beta u_\beta(0)}\epsilon^\alpha\\\nonumber
    &&+\frac{q n^\alpha}{m (n^\beta u_\beta(0))^2}\Big(2a\epsilon^\beta u_\beta(0)\sin^2(\zeta/2)\\\nonumber
    &&+\frac{q a^2}{4m}(\zeta-\frac{1}{2}\sin(2\zeta))\Big)
\end{eqnarray}
with $\zeta=n^\alpha u_\alpha(0)\tau$; see for instance \cite{itzykson2012quantum}. 
Hence, on average, the plane wave transports a particle along with it with the constant velocity
\begin{equation}
   \bar{u}^\alpha(\tau)= \frac{q^2a^2}{4m^2(n^\alpha u_\alpha(0))}\,.
\end{equation}
Given an initial velocity opposite to this transport velocity
\begin{equation}\label{vinit}
    n_i u^i(0)=\frac{-\frac{q^2a^2}{4m^2}}{\sqrt{1+\frac{q^2a^2}{2m^2}}},
    \qquad
     i=1,2,3\,,
\end{equation}
the particle only oscillates with the famous figure 8 motion. 

\begin{figure}
  \includegraphics[width=\linewidth]{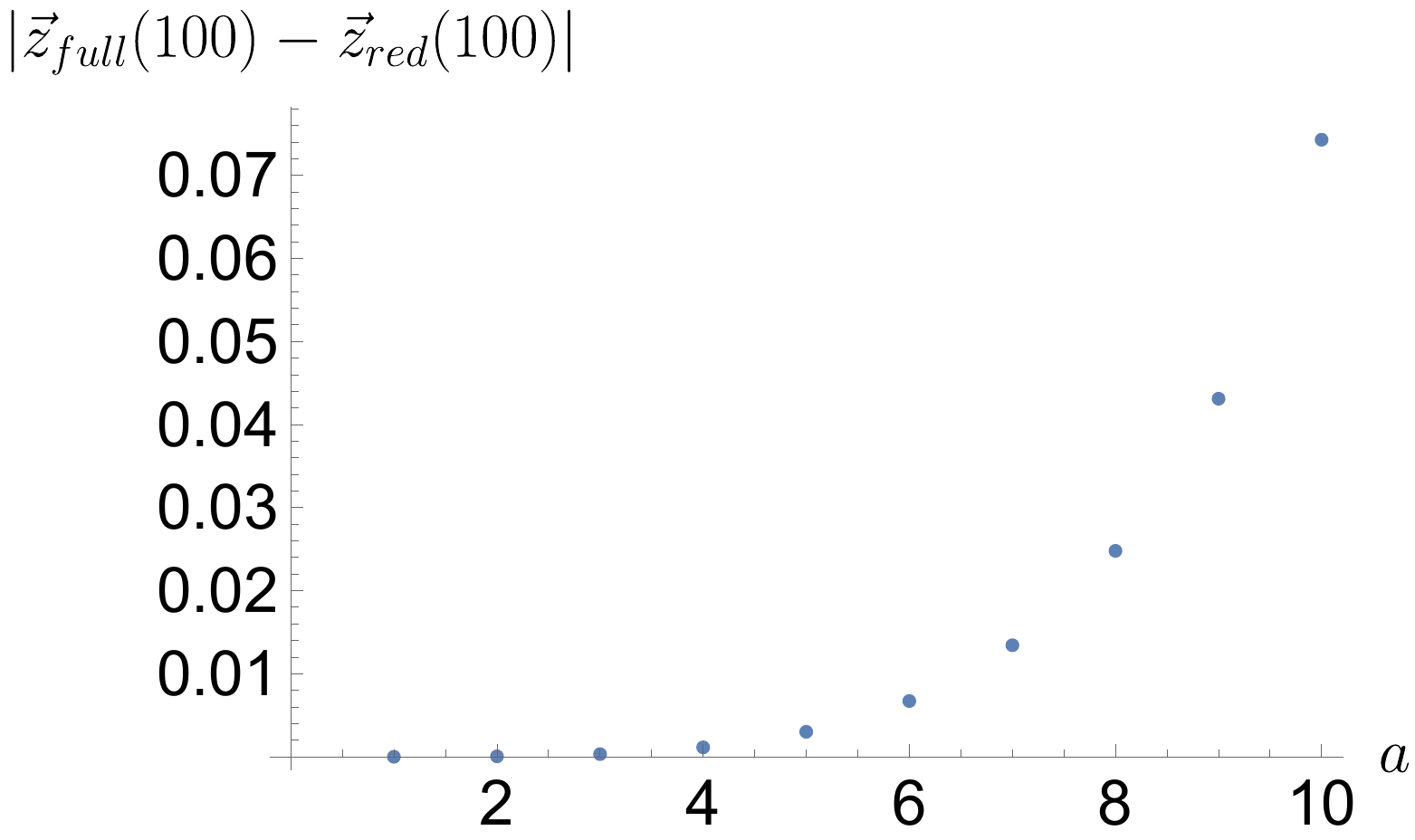}
  \caption{The figure shows the absolute value for the difference in the 3d-positions of the full DSR and the reduced DSR equations for different values of  $a$ after a duration of 100 in a plane wave.}
  \label{plane_wave_full_red_div}
\end{figure}

\begin{figure}
  \includegraphics[width=\linewidth]{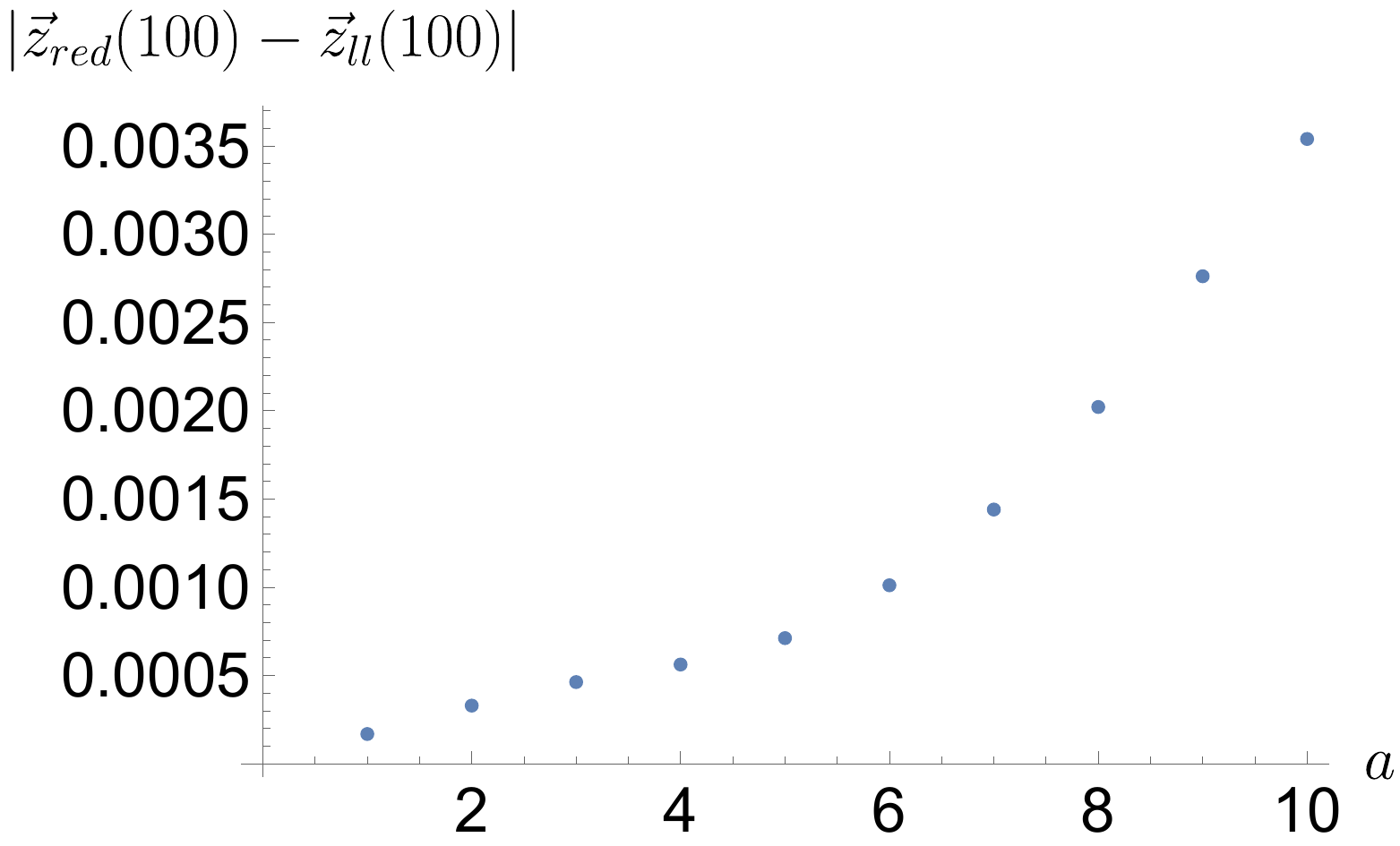}
  \caption{The figure shows the absolute value for the difference in the 3d-positions of the LL equation and the reduced DSR equation for different values of  $a$ after a duration of 100 in a plane wave.}
  \label{plane_wave_ll_red_div}
\end{figure}

When radiation reaction comes into the picture, the particle is accelerated and the originally stable figure 8 motion is altered and, on average, accelerated in the direction of the pulse. Figure~\ref{8red} shows the resulting acceleration due to the radiation reaction for 20 units of time and $a=1$. The fact that the radiation reaction force accelerates the particle instead of slowing it down is not in contradiction with energy conservation. The emitted radiation of the particle interferes destructively with the plane wave, and thus, the absorption and not emission of energy needs to be compensated by the radiation reaction force. An explicit solution for the LL equation is also available \cite{di2008exact,hadad2010effects} but in present paper numerical solutions are used.
We compare the 3d-position after a duration of 100 for the field strength a between 1 and 10 with an initial velocity given by \eqref{vinit} and $\epsilon=0.05$ as before. The absolute value of the difference
of the 3d-positions $\vec{z}(\tau)$ are shown in Figure \ref{plane_wave_full_red_div} and Figure \ref{plane_wave_ll_red_div}.
For $a=10$ the particle has been transported for 23 distance units in z direction and reached a velocity of 0.68c. In this case the reduced DSR equation show much closer agreement with the LL equation. The deviations for the full and the reduced DSR equation are almost identical to the deviations between the full DSR and the LL equations, hence they are not shown. The deviations are bigger than in case a) of the constant magnetic field, but it has to be emphasized that the magnetic field is confining, and therefore, the covered distance is much smaller. In proportion to the covered distance, the deviations are comparable in both cases.



\subsection{Main Result II: The $\epsilon$-dependence of the solution to the DSR equations}
\label{sec_epsilon_dep}

Since an explicit solution is only available for the case of a constant electric field and since in this case the $\epsilon$-dependence of the solutions to the DSR equations vanishes after renormalization, so far only a numerical analysis allows to study the $\epsilon$-dependence of the DSR equations. 

\begin{figure}
  \includegraphics[width=\linewidth]{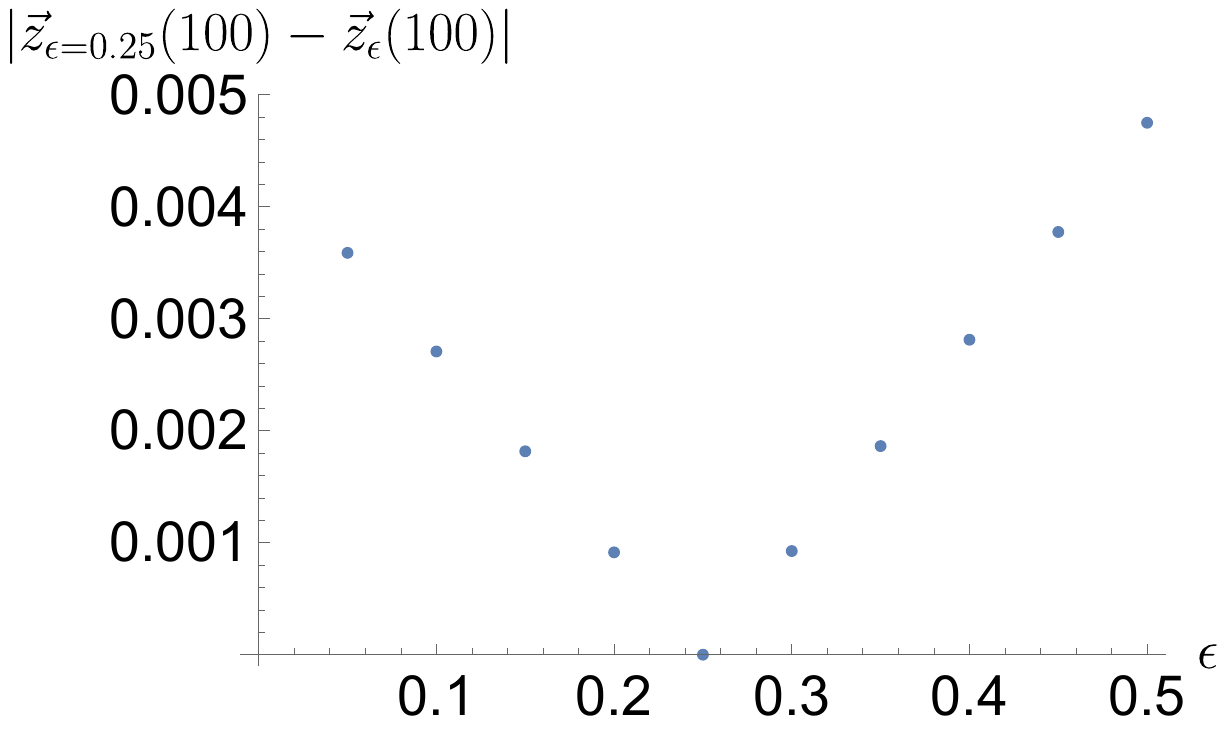}
  \caption{The figure shows the absolute value for the difference in the 3d-positions for different values of  $\epsilon$ after a duration of 100 in a constant magnetic field of strength 10 compared to the solution with  $\epsilon=0.25$ for the full DSR equation.}
  \label{epsi_dep_full}
\end{figure}

\begin{figure}
  \includegraphics[width=\linewidth]{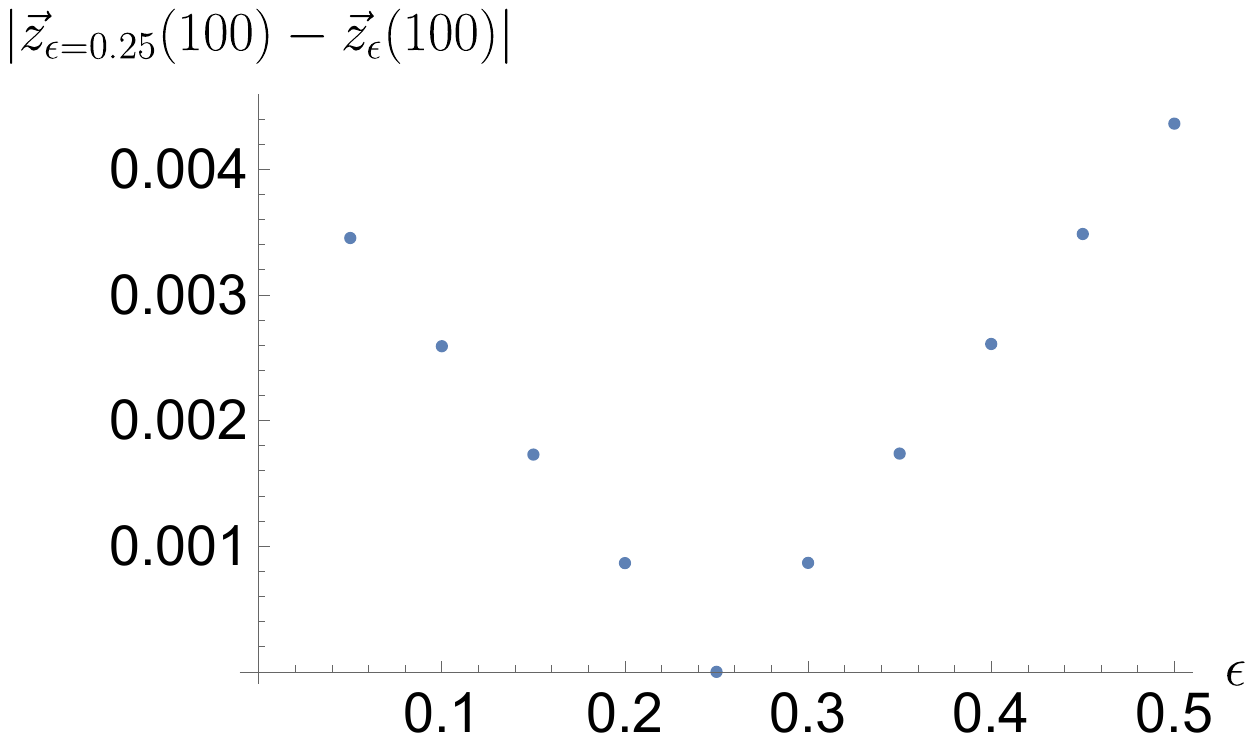}
  \caption{The figure shows the absolute value for the difference in the 3d-positions for different values of  $\epsilon$ after a duration of 100 in a constant magnetic field of strength 10 compared to the solution with  $\epsilon=0.25$ for the reduced DSR equation.}
  \label{epsi_dep_red}
\end{figure}

For this purpose, we revisit the constant magnetic field setting of Section \ref{sec_comp_LL} in case a) with the same magnetic field of strength 10 and initial velocity of 0.1. After the duration of 100 time units we compare solutions for 10 different values of $\epsilon$ ranging between $0.05$ and $0.5$ with a uniform step size of $0.05$. The difference in 3d-position w.r.t.\ the case of $\epsilon=0.25$ are shown in Figure~\ref{epsi_dep_full}. These show an approximately linear dependence in $\epsilon$ with a slope of $10^{-2}$. This is expected since there is no term of the order $\epsilon^{-1}$ thanks to the mass renormalization and higher order contributions are small. The smallness of the $\epsilon$-dependence of the solution is surprising since the radius of the initial orbit is of 0.1 distance units. Accordingly, $\epsilon=0.25$ corresponds to almost half of a rotation. In general, the independence of $\epsilon$ can only be expected, if $\epsilon$ is smaller than the curvature of the trajectory, but nevertheless the deviation is still small in this situation. For weaker fields, the $\epsilon$-dependence is much smaller since the curvature radius of the trajectory is larger and the radiation reaction force is smaller in total. For the reduced DSR equation, the distances are shown in Figure~\ref{epsi_dep_red}. There is also a linear dependence on $\epsilon$ with a similar slope. For weaker fields, the full DSR equation shows a smaller dependency than the reduced DSR equation.

As this numerical example shows, there seems to be only a weak dependence of the solutions on $\epsilon$ in the considered regimes -- and this also was our general impression gained from multiple other numerical setups in our investigation. We therefore believe that this is a general feature of both the full DSR and reduced DSR equations. A general proof of this statement, of course, would be desirable but is still open. Interestingly, the numerical solver becomes unstable when $\epsilon$ approaches the classical electron radius even in the case of uniform acceleration, which is likely due to the numerical instability discussed in Section~\ref{sec_flow}.

\section{Summary and Outlook}
\label{sec_outlook}

We have reviewed a recent proposal of the full DSR equation \eqref{DSR} and presented a new reduced DSR force \eqref{red-flow} in Sections \ref{sec_dsr_force} and \ref{sec_reduced_model}. In Sections \ref{sec_flow} and
\ref{sec_numerical_method} we have discussed the numerical solver of the full DSR
and the reduced DSR equations. In Section \ref{sec_remark_uniqueness} we have analyzed
the existence of solutions of the full DSR equation. In Section
\ref{sec_units} the units used for simulations have been presented and in Section
\ref{sec_comp_LL} we have shown the physical plausibility of both DSR equations
in three settings relevant to radiation reaction phenomena. For this purpose, the trajectories of both DSR
dynamics have been compared to the trajectories of the LL dynamics, c.f.\ Section
\ref{sec_LL_dynamics}) below, in the case of a constant electric field, a constant
magnetic field, and a plane wave. In all three cases, all three models
have agreed with high precision. Given the different structures of the three
models, in particular the different types of delay and non-delay equations, it was not clear a priory that both DSR models would pass this test. On
the other hand, the small differences suggest that it is challenging to
distinguish the three models experimentally. For regimes in which this might be
possible, possibly already quantum corrections may have to be taken into account. Such quantum
corrections are planed to be investigated in future work. Section \ref{sec_epsilon_dep}
suggests that the $\epsilon$-dependence is weak and does not lead to practical problems as long as the discussed singularity marked by the classical electron radius is avoided.
From a practical point of view, both the full DSR and especially the reduced
DSR equations are computationally considerably less costly in  multi-particle simulations
as elaborated in Section \ref{sec_efficiency} below.

\section{Acknowledgement}

C.B. and H.R. acknowledge the hospitality of the
Arnold Sommerfeld Center at the Ludwig-Maximilians-
Universit\"at in Munich. This work has been funded by the
Deutsche Forschungsgemeinschaft (DFG) under Grant
No. 416699545 within the Research Unit FOR2783/1.
This work was partly supported by the Elite Network of Bavaria through the Junior Research Group ``Interaction between Light and Matter''.

\appendix

\section{The Landau-Lifschitz equation}
\label{sec_LL_dynamics}

For the reason of self-containedness, we give
a brief introduction to the LL system of equations that is used as a
reference system. With regards to the rigorous derivation
of the LL equation, we refer the reader to \cite{spohn2004dynamics}. For our
purposes here, an informal motivation of the LL equation suffices, which is
usually done by means of a perturbation argument of the LAD equation. The
radiation reaction force is then treated as a small correction only to first
order. Such a first order approximation may be reasonable for a wide range of
parameters but within strong and rapidly changing fields, however, the radiation reaction force
might become the dominant force. To infer the LL equation in such a fashion,
every appearance of the acceleration within in the LAD equation is simply
replaced by the Lorentz force, which results in:
\begin{eqnarray}\label{lleqm}
    ma^\alpha&=&qF^{\alpha\beta}u_\beta+\frac{2q^2}{3}\Big(\frac{d}{d\tau}a^\alpha+a^\gamma a_\gamma u^\alpha\Big)\nonumber\\
    &\approx&  qF^{\alpha\beta}u_\beta+\frac{2q^2}{3}\Big(\frac{q}{m}\frac{d}{d\tau}\Big(F^{\alpha\beta}u_\beta\Big)\nonumber\\
    &&+\frac{q^2}{m^2}F^{\gamma\beta}u_\beta F_{\gamma\mu}u^\mu u^\alpha\Big)\nonumber\\
    &=& qF^{\alpha\beta}u_\beta+\frac{2q^2}{3}\Big(\frac{q}{m}\Big(\frac{\partial F^{\alpha\beta}}{\partial x^\nu}u^\nu u_\beta+F^{\alpha\beta}a_\beta\Big)\nonumber\\
    &&+\frac{q^2}{m^2}F^{\gamma\beta}u_\beta F_{\gamma\mu}u^\mu u^\alpha\Big)\nonumber\\
    &\approx& qF^{\alpha\beta}u_\beta+\frac{2q^2}{3}\Big(\frac{q}{m}\frac{\partial F^{\alpha\beta}}{\partial x^\nu}u^\nu u_\beta+ \frac{q^2}{m^2}F^{\alpha\beta}F_{\beta\kappa}u^\kappa\nonumber\\
    &&+\frac{q^2}{m^2}F^{\gamma\beta}u_\beta F_{\gamma\mu}u^\mu u^\alpha\Big)\,.
\end{eqnarray}
We have used a similar numerical integrator as discussed in
Section~\ref{sec_numerical_method} in our investigations. In this regard, we note that the
Gauss-Legendre methods conserve quadratic invariants. Here, the norm $u^\alpha
u_\alpha=1$ is conserved with the same precision as the orthogonality of
$a^\alpha u_\alpha=0$. Equation \eqref{lleqm} is numerically not
perfectly suited to ensure $a^\alpha u_\alpha=0$ since it leads to subtractions
between terms of similar size. Instead, the form 
\begin{equation}
    ma^\alpha=qF^{\alpha\beta}u_\beta+\frac{2q^2}{3}(v^\alpha-v^\beta u_\beta u^\alpha)
\end{equation}
with 
\begin{equation}
    v^\alpha=\frac{q}{m}\frac{\partial F^{\alpha\beta}}{\partial x^\nu}u^\nu u_\beta+ \frac{q^2}{m^2}F^{\alpha\beta}F_{\beta\kappa}u^\kappa
\end{equation}
ensures the orthogonality to machine precision. This is also discussed in \cite{elkina2014improving}.

\subsection{Tests of the numerical solver}
\label{sec_tests}

The numerical solver described in Section~\ref{sec_numerical_method} is
employed for the DSR, reduced DSR, and the LL equations alike. In the following, we benchmark it against explicitly know solutions.

\begin{figure}
  \includegraphics[width=\linewidth]{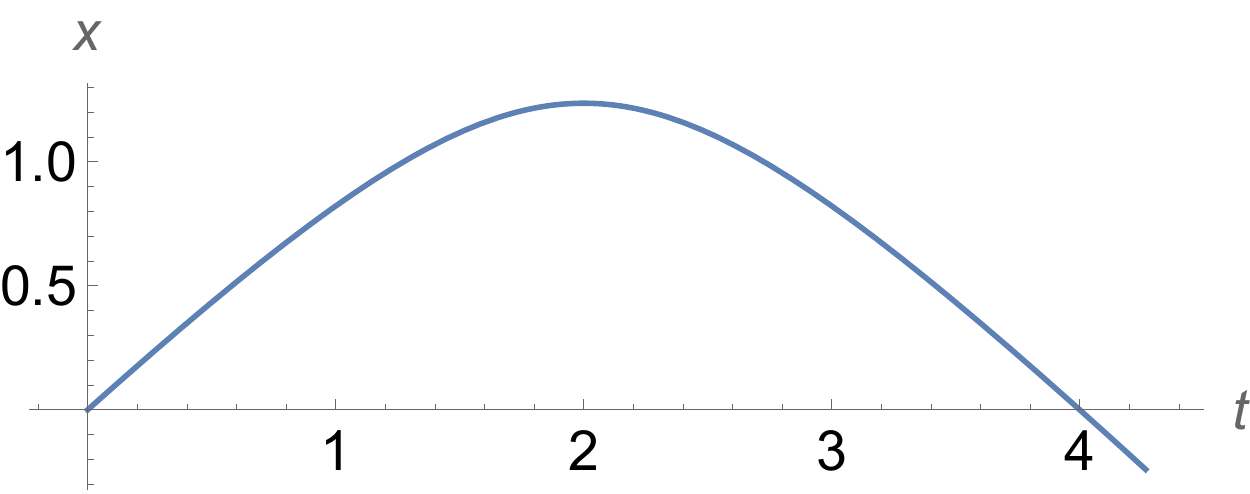}
  \caption{The figure shows the trajectory of a particle moving against an uniform electric field for the eigentime duration 3.}
  \label{ana_sol_const}
\end{figure}

\begin{figure}
  \includegraphics[width=\linewidth]{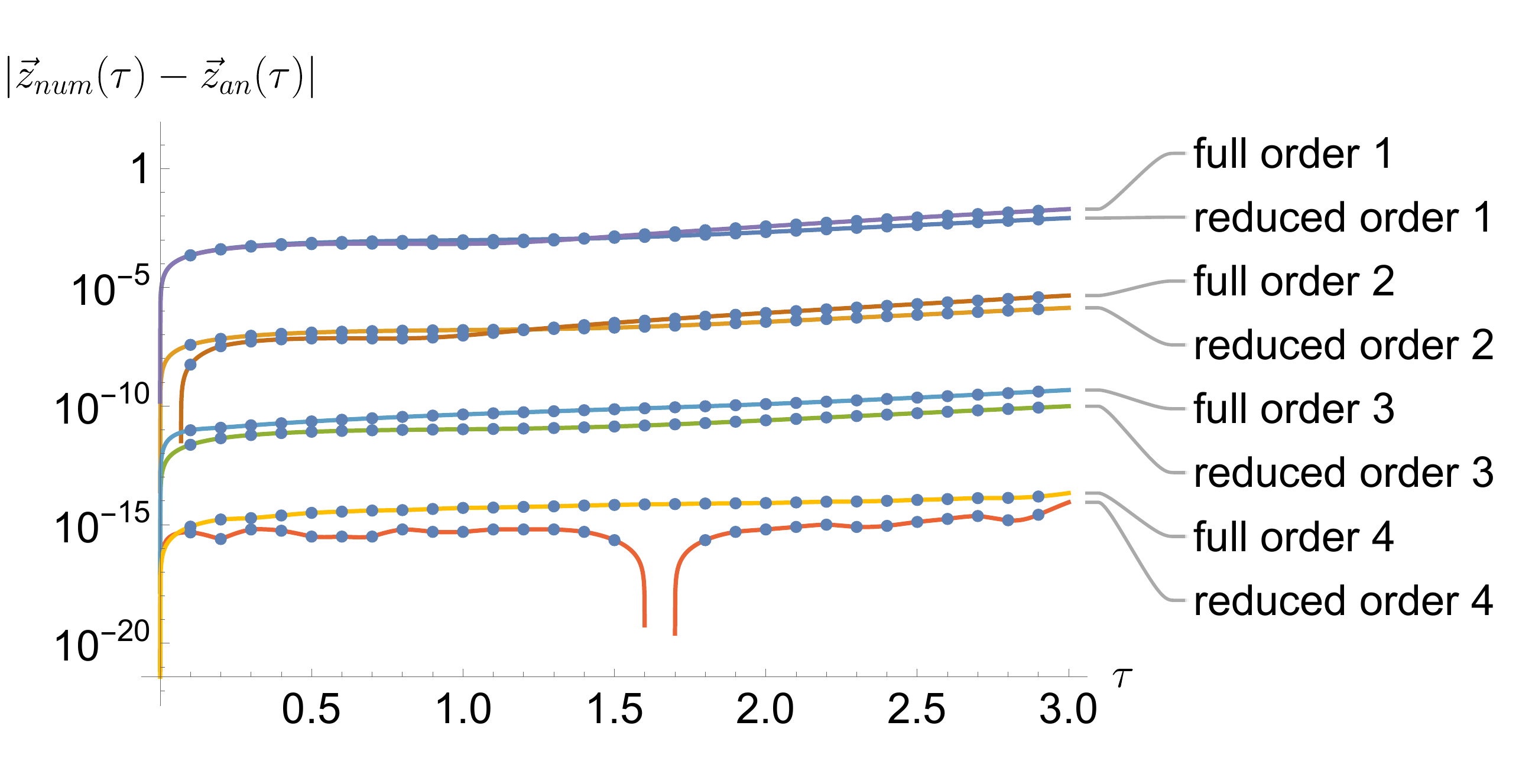}
  \caption{The figure shows the deviation of the numerical solution from the
    explicit solution for the full DSR and the reduced DSR equations for the trajectory of Figure~\ref{ana_sol_const}. The downwards kink in the computation of order 4 is due to a coincidence in which the numerical error was exactly zero.}
  \label{errors_const_acc}
\end{figure}

\paragraph{DSR and reduced DSR for constant electric fields:}
An
explicit solution is yet only available for the full DSR and reduced DSR equations in the
case of uniform acceleration,
\begin{equation}\label{constacc}
	z^\alpha(\tau)=\frac{1}{g}(\sinh(g\tau),0,0,\cosh(g\tau))\,,
\end{equation}
where $g$ is a constant acceleration in a purely electric field along the
z-direction. In this case, the radiation reaction force contributes only in the
form of mass renormalization. Inserting \eqref{constacc} into the full DSR
equation \eqref{endgl} leads to 
\begin{equation}
	m_\text{eff}=m+\frac{q^2g}{2\sinh(g\epsilon)}
\end{equation}
 and for the reduced DSR equation \eqref{red-flow}, for which the mass has already been
 renormalized to lowest order, one obtains
\begin{equation}\label{eq_entire_inertia}
	m_\text{eff}=m+\frac{4q^2}{3\epsilon}\sinh(\frac{g\epsilon}{2})^2\,.
\end{equation}
As a test case, the initial conditions $u_x(0)=2$ and zero for the y and z components are chosen, the mass is set to $m=1$, the delay to $\epsilon=0.1$, the step size to $h=0.1$, the charge to $q=0.1$, the magnetic field to zero, and the electric field to $E_z=-10\, m_\text{eff}$ corresponding to an acceleration $g=10$. The solution is shown in Figure~\ref{ana_sol_const}. 
In Figure~\ref{errors_const_acc}, the differences between the explicit \eqref{constacc} and numerical solutions for the full DSR and the reduced DSR equations are shown. The numerical error of the full DSR solution is slightly above the error of the reduced solution for all orders shown. Since the computational effort for the full DSR equation is considerably larger than for the reduced DSR equation, this behavior is expected. Already for the fourth order Gauss-Legendre method, the numerical error is of the order of the machine precision.   
For uniform acceleration the radiation reaction force due to the LL equation (and also due to the LAD equation) is zero. Hence, this case is not suitable for testing the solver for the LL equation.\\
\begin{figure}
  \includegraphics[width=\linewidth]{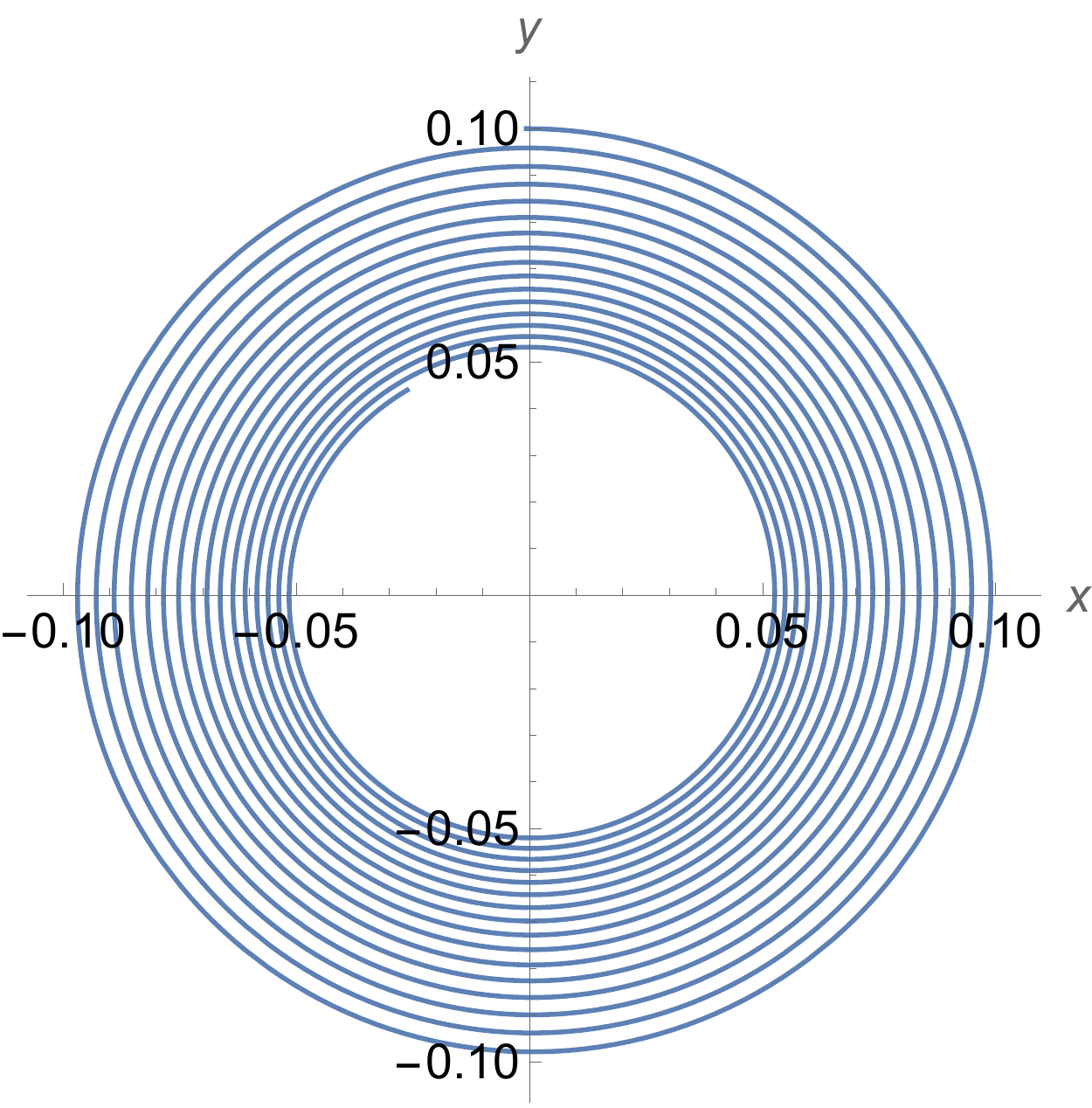}
  \caption{The figure shows the x and y components of the explicit solutions of the LL equation in a constant magnetic field of strength 10 in z direction with initial four velocity $u^2(0)=0.1$.}
  \label{const_mag_ana_ll}
\end{figure}

\begin{figure}
  \includegraphics[width=\linewidth]{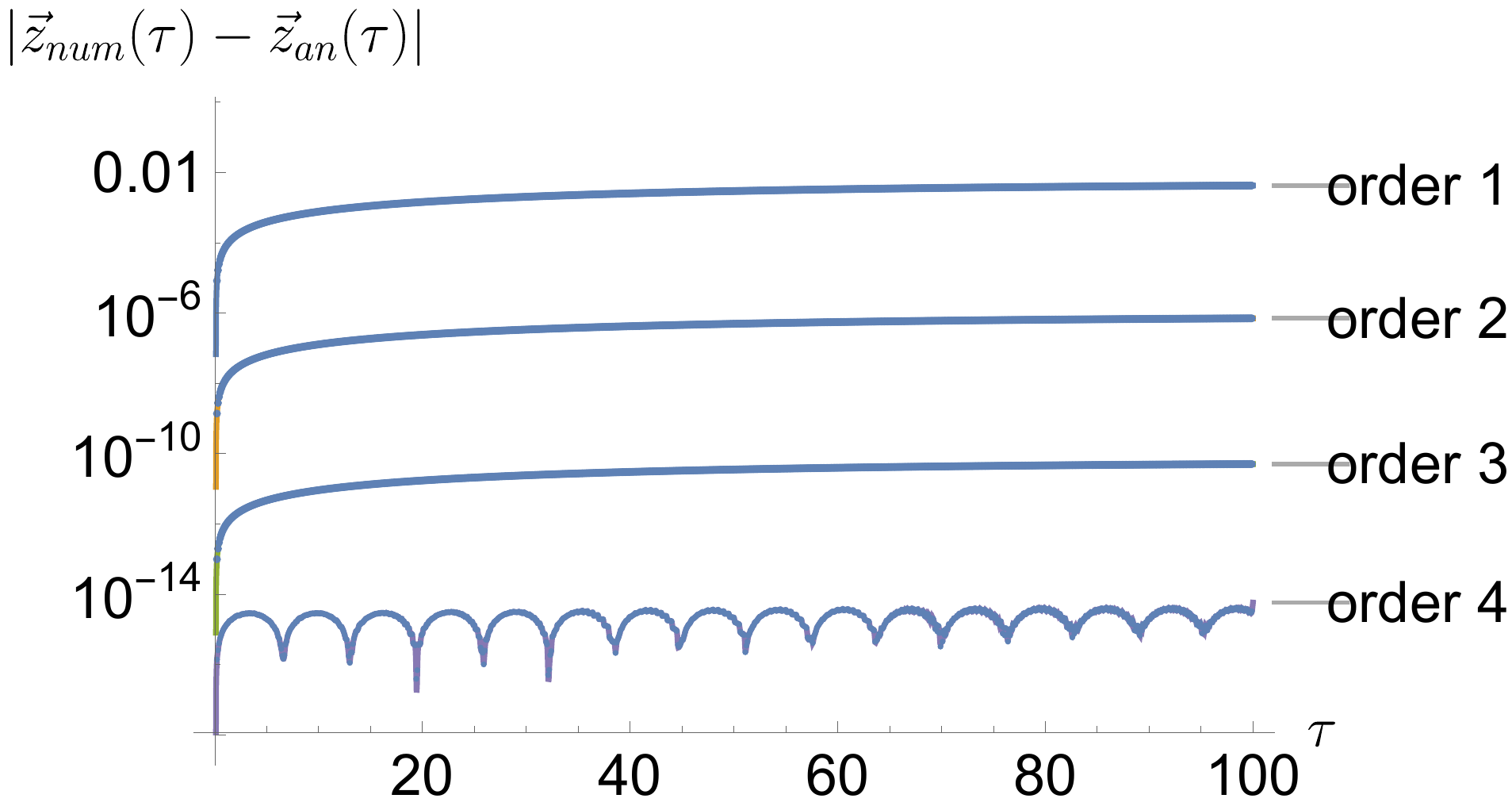}
  \caption{The figure shows the error between the numerical and the explicit solution for the LL equation for the trajectory of Figure~\ref{const_mag_ana_ll}}
  \label{errors_const_mag}.
\end{figure}

\paragraph{LL for constant magnetic fields:}
 Another example for testing is a particle in a constant magnetic field. In
 this case an explicit solution is available only for the LL equation. The derivation can be found in \cite{elkina2014improving}. The solution is given by
\begin{eqnarray}\label{ll-const-mag}
&&\begin{pmatrix}
		x-x_0\\y-y_0
	\end{pmatrix}=\nonumber\\
	&&\frac{\sigma}{u_0}(\sqrt{(1+u_0^2)e^{2K\phi}-u_0^2}\begin{pmatrix}
		Im\\-Re
	\end{pmatrix}
		e^{\phi_0-\sigma \phi}\nonumber\\
		&&\times_2F_1(1,\frac{1}{2}-\frac{i\sigma}{2K},1-\frac{i\sigma}{2K},\frac{1+u_0^2}{u_0^2}e^{2K\phi})\Big\rvert^\phi_0
\end{eqnarray}

with
\begin{eqnarray}
K=\frac{2q^3B}{3m^2}\,,\qquad\phi=\frac{eB\tau}{m}\,,\\ 
u_z=0\,, \quad u_0^2=u_x^2(0)+u_y^2(0)
\end{eqnarray}
and $\sigma$ is the sign of the charge. Figure \ref{const_mag_ana_ll} shows the
trajectory and Figure~\ref{errors_const_mag} the difference between the
numerical and explicit solutions for different orders of the solver. The parameters $q=0.1$, $m=1$, $u_0=(\sqrt{1.01},0.1,0,0)$, $z_0=(0,0,0.1,0)$, $\vec{B}=10\Vec{e}_z$, and the proper time duration of $\tau=100$ have been chosen.

\subsection{Numerical efficiency}
\label{sec_efficiency}

To test the numerical performance of the three different equations of motion,
i.e., the full DSR, reduced DSR, and LL equations, we measure the time needed for the simulations.
First, we consider one particle in an external field. While there are some
fluctuations for the execution time for the different equations and the used
numerical integration orders, overall the LL integrator is a bit faster than
the ones of the full DSR equation for constant fields. But with coordinate dependent
fields, the integration of the full DSR equation appeared on average 3 times faster than the LL equation while the one of the
reduced DSR equation appeared on average 4 times faster than the one of the full DSR
equation. Second, for multi-particle simulations, it can be expected that the
performance of the LL integrator is significantly worse, since the derivative of
the field strength tensor is needed. The field strength tensor generated by one
particle is given by
\begin{eqnarray}
F^{\alpha\beta}&=&\frac{q}{r^2}\left( w^\alpha u^\beta-u^\alpha w^\beta\right)\nonumber\\
&&+\frac{q}{r}\big((u^\alpha+w^\alpha)(a^\beta+a^\gamma w_\gamma w^\beta)\nonumber\\
&&-(u^\beta+w^\beta)(a^\alpha+a^\gamma w_\gamma w^\alpha)\big)
\end{eqnarray}
with $r=(x^\alpha-z^\alpha)u_\alpha$ and $w^\alpha=(x^\alpha-z^\alpha)/r-u^\alpha$. Positions $z^\alpha$, velocities $u^\alpha$, and the accelerations $a^\alpha$ are needed at the retarded time. With the help of 
\begin{equation}
    \frac{\partial \tau_{ret}}{\partial x_\alpha}=u^\alpha+w^\alpha
\end{equation}
the derivative of the field strength tensor is given by
\begin{eqnarray}\label{fieldder}
&&\frac{\partial F^{\alpha\beta}}{\partial x_\gamma}=\nonumber\\
&&\frac{1}{r^3}
    \eta^{\alpha\gamma} u^\beta  - \frac{1}{r^3} \eta^{\beta\gamma} u^\alpha  - \frac{3}{r^3} u^\alpha w^\beta w^\gamma \nonumber\\
&&  + \frac{3}{r^3} u^\beta w^\alpha w^\gamma  - \frac{1}{r^2} a^\alpha \eta^{\beta\gamma}  - \frac{3}{r^2} a^\alpha u^\beta w^\gamma \nonumber\\
&&  - \frac{1}{r^2} a^\alpha u^\gamma w^\beta  - \frac{3}{r^2} a^\alpha w^\beta
    w^\gamma  + \frac{1}{r^2} a^\beta \eta^{\alpha\gamma} \nonumber\\
&&  + \frac{3}{r^2} a^\beta u^\alpha w^\gamma  + \frac{1}{r^2} a^\beta u^\gamma w^\alpha  + \frac{3}{r^2} a^\beta w^\alpha w^\gamma \nonumber\\
&&  + \frac{1}{r^2} a^\gamma u^\alpha w^\beta  - \frac{1}{r^2} a^\gamma u^\beta w^\alpha  + \frac{3}{r^2} a^\delta u^\alpha u^\gamma w^\beta w_\delta \nonumber\\
&&  + \frac{3}{r^2} a^\delta u^\alpha w^\beta w^\gamma w_\delta  - \frac{3}{r^2} a^\delta u^\beta u^\gamma w^\alpha w_\delta  - \frac{3}{r^2} a^\delta u^\beta w^\alpha w^\gamma w_\delta \nonumber\\    
&&  - \frac{1}{r} a^\alpha a^\gamma u^\beta  - \frac{1}{r} a^\alpha a^\gamma w^\beta  + \frac{2}{r} a^\alpha a^\delta u^\beta u^\gamma w_\delta \nonumber\\
&&  + \frac{2}{r} a^\alpha a^\delta u^\beta w^\gamma w_\delta  + \frac{2}{r} a^\alpha a^\delta u^\gamma w^\beta w_\delta  + \frac{2}{r} a^\alpha a^\delta w^\beta w^\gamma w_\delta \nonumber\\
&&  + \frac{1}{r} a^\beta a^\gamma u^\alpha  + \frac{1}{r} a^\beta a^\gamma w^\alpha  - \frac{2}{r} a^\beta a^\delta u^\alpha u^\gamma w_\delta \nonumber\\
&&  - \frac{2}{r} a^\beta a^\delta u^\alpha w^\gamma w_\delta  - \frac{2}{r} a^\beta a^\delta u^\gamma w^\alpha w_\delta  - \frac{2}{r} a^\beta a^\delta w^\alpha w^\gamma w_\delta \nonumber\\
&&  - \frac{1}{r} a^\gamma a_\gamma u^\alpha w^\beta w_\gamma  + \frac{1}{r} a^\gamma a_\gamma u^\beta w^\alpha w_\gamma  + \frac{3}{r} a^\gamma a^\delta u^\alpha w^\beta w_\delta \nonumber\\
&&  - \frac{3}{r} a^\gamma a^\delta u^\beta w^\alpha w_\delta  - \frac{1}{r} \dot{a}^\alpha u^\beta u^\gamma  - \frac{1}{r} \dot{a}^\alpha u^\beta w^\gamma \nonumber\\
&&  - \frac{1}{r} \dot{a}^\alpha u^\gamma w^\beta  - \frac{1}{r} \dot{a}^\alpha w^\beta w^\gamma  + \frac{2}{r} \dot{a}^\beta u^\alpha u^\gamma \nonumber\\
&&  + \frac{1}{r} \dot{a}^\beta u^\alpha w^\gamma  + \frac{1}{r} \dot{a}^\beta u^\gamma w^\alpha  + \frac{1}{r} \dot{a}^\beta w^\alpha w^\gamma \nonumber\\
&&  - \frac{1}{r} \dot{a}^\gamma u^\alpha w^\beta  + \frac{1}{r} \dot{a}^\gamma u^\beta w^\alpha\,.
\end{eqnarray}
Since this is a lengthy expression, which must be evaluated for each respectively other particle the numerical cost of the LL integrator is much higher than the cost for both DSR integrators. A way to avoid (\ref{fieldder}) is by making use of numerical derivatives. Even though there are still 24 different derivatives needed, this gives a considerable speed improvement. The disadvantage of numerical derivatives is a significant precision loss. This loss appears because two numbers of almost equal magnitude have to be subtracted from each other. Hence, usually half of the significant digits are lost, which makes this option unattractive for high precision simulations.

\bibliography{references.bib}

\begin{thebibliography}{12}%
\makeatletter
\providecommand \@ifxundefined [1]{%
 \@ifx{#1\undefined}
}%
\providecommand \@ifnum [1]{%
 \ifnum #1\expandafter \@firstoftwo
 \else \expandafter \@secondoftwo
 \fi
}%
\providecommand \@ifx [1]{%
 \ifx #1\expandafter \@firstoftwo
 \else \expandafter \@secondoftwo
 \fi
}%
\providecommand \natexlab [1]{#1}%
\providecommand \enquote  [1]{``#1''}%
\providecommand \bibnamefont  [1]{#1}%
\providecommand \bibfnamefont [1]{#1}%
\providecommand \citenamefont [1]{#1}%
\providecommand \href@noop [0]{\@secondoftwo}%
\providecommand \href [0]{\begingroup \@sanitize@url \@href}%
\providecommand \@href[1]{\@@startlink{#1}\@@href}%
\providecommand \@@href[1]{\endgroup#1\@@endlink}%
\providecommand \@sanitize@url [0]{\catcode `\\12\catcode `\$12\catcode
  `\&12\catcode `\#12\catcode `\^12\catcode `\_12\catcode `\%12\relax}%
\providecommand \@@startlink[1]{}%
\providecommand \@@endlink[0]{}%
\providecommand \url  [0]{\begingroup\@sanitize@url \@url }%
\providecommand \@url [1]{\endgroup\@href {#1}{\urlprefix }}%
\providecommand \urlprefix  [0]{URL }%
\providecommand \Eprint [0]{\href }%
\providecommand \doibase [0]{https://doi.org/}%
\providecommand \selectlanguage [0]{\@gobble}%
\providecommand \bibinfo  [0]{\@secondoftwo}%
\providecommand \bibfield  [0]{\@secondoftwo}%
\providecommand \translation [1]{[#1]}%
\providecommand \BibitemOpen [0]{}%
\providecommand \bibitemStop [0]{}%
\providecommand \bibitemNoStop [0]{.\EOS\space}%
\providecommand \EOS [0]{\spacefactor3000\relax}%
\providecommand \BibitemShut  [1]{\csname bibitem#1\endcsname}%
\let\auto@bib@innerbib\@empty
\bibitem [{\citenamefont {Bild}\ \emph {et~al.}(2019)\citenamefont {Bild},
  \citenamefont {Deckert},\ and\ \citenamefont {Ruhl}}]{bild2019radiation}%
  \BibitemOpen
  \bibfield  {author} {\bibinfo {author} {\bibfnamefont {C.}~\bibnamefont
  {Bild}}, \bibinfo {author} {\bibfnamefont {D.-A.}\ \bibnamefont {Deckert}},\
  and\ \bibinfo {author} {\bibfnamefont {H.}~\bibnamefont {Ruhl}},\ }\bibfield
  {title} {\bibinfo {title} {Radiation reaction in classical electrodynamics},\
  }\href@noop {} {\bibfield  {journal} {\bibinfo  {journal} {Physical Review
  D}\ }\textbf {\bibinfo {volume} {99}},\ \bibinfo {pages} {096001} (\bibinfo
  {year} {2019})}\BibitemShut {NoStop}%
\bibitem [{\citenamefont {Dirac}(1938)}]{dirac1938classical}%
  \BibitemOpen
  \bibfield  {author} {\bibinfo {author} {\bibfnamefont {P.~A.}\ \bibnamefont
  {Dirac}},\ }\bibfield  {title} {\bibinfo {title} {{Classical theory of
  radiating electrons}},\ }\href@noop {} {\bibfield  {journal} {\bibinfo
  {journal} {Proceedings of the Royal Society of London. Series A, Mathematical
  and Physical Sciences}\ ,\ \bibinfo {pages} {148}} (\bibinfo {year}
  {1938})}\BibitemShut {NoStop}%
\bibitem [{\citenamefont {Spohn}(2004)}]{spohn2004dynamics}%
  \BibitemOpen
  \bibfield  {author} {\bibinfo {author} {\bibfnamefont {H.}~\bibnamefont
  {Spohn}},\ }\href@noop {} {\emph {\bibinfo {title} {Dynamics of charged
  particles and their radiation field}}}\ (\bibinfo  {publisher} {Cambridge
  university press},\ \bibinfo {year} {2004})\BibitemShut {NoStop}%
\bibitem [{\citenamefont {Bauer}\ and\ \citenamefont
  {D{\"u}rr}(2001)}]{bauer2001}%
  \BibitemOpen
  \bibfield  {author} {\bibinfo {author} {\bibfnamefont {G.}~\bibnamefont
  {Bauer}}\ and\ \bibinfo {author} {\bibfnamefont {D.}~\bibnamefont
  {D{\"u}rr}},\ }\bibfield  {title} {\bibinfo {title} {The maxwell-lorentz
  system of a rigid charge},\ }\href@noop {} {\bibfield  {journal} {\bibinfo
  {journal} {Annales Henri Poincar{\'e}}\ }\textbf {\bibinfo {volume} {2}},\
  \bibinfo {pages} {179} (\bibinfo {year} {2001})}\BibitemShut {NoStop}%
\bibitem [{\citenamefont {Faci}\ and\ \citenamefont
  {Novello}(2016)}]{faci2016time}%
  \BibitemOpen
  \bibfield  {author} {\bibinfo {author} {\bibfnamefont {S.}~\bibnamefont
  {Faci}}\ and\ \bibinfo {author} {\bibfnamefont {M.}~\bibnamefont {Novello}},\
  }\bibfield  {title} {\bibinfo {title} {Time-delayed electromagnetic radiation
  reaction},\ }\href@noop {} {\bibfield  {journal} {\bibinfo  {journal} {arXiv
  preprint arXiv:1611.07611}\ } (\bibinfo {year} {2016})}\BibitemShut {NoStop}%
\bibitem [{\citenamefont {Hairer}\ \emph {et~al.}(2006)\citenamefont {Hairer},
  \citenamefont {Lubich},\ and\ \citenamefont {Wanner}}]{hairer2006geometric}%
  \BibitemOpen
  \bibfield  {author} {\bibinfo {author} {\bibfnamefont {E.}~\bibnamefont
  {Hairer}}, \bibinfo {author} {\bibfnamefont {C.}~\bibnamefont {Lubich}},\
  and\ \bibinfo {author} {\bibfnamefont {G.}~\bibnamefont {Wanner}},\
  }\href@noop {} {\bibinfo {title} {Geometric numerical integration, volume 31
  of springer series in computational mathematics}} (\bibinfo {year}
  {2006})\BibitemShut {NoStop}%
\bibitem [{\citenamefont {Iserles}(2009)}]{iserles2009first}%
  \BibitemOpen
  \bibfield  {author} {\bibinfo {author} {\bibfnamefont {A.}~\bibnamefont
  {Iserles}},\ }\href@noop {} {\emph {\bibinfo {title} {A first course in the
  numerical analysis of differential equations}}},\ \bibinfo {number} {44}\
  (\bibinfo  {publisher} {Cambridge university press},\ \bibinfo {year}
  {2009})\BibitemShut {NoStop}%
\bibitem [{\citenamefont {Driver}(2012)}]{driver2012ordinary}%
  \BibitemOpen
  \bibfield  {author} {\bibinfo {author} {\bibfnamefont {R.~D.}\ \bibnamefont
  {Driver}},\ }\href@noop {} {\emph {\bibinfo {title} {Ordinary and delay
  differential equations}}},\ Vol.~\bibinfo {volume} {20}\ (\bibinfo
  {publisher} {Springer Science \& Business Media},\ \bibinfo {year}
  {2012})\BibitemShut {NoStop}%
\bibitem [{\citenamefont {Itzykson}\ and\ \citenamefont
  {Zuber}(2012)}]{itzykson2012quantum}%
  \BibitemOpen
  \bibfield  {author} {\bibinfo {author} {\bibfnamefont {C.}~\bibnamefont
  {Itzykson}}\ and\ \bibinfo {author} {\bibfnamefont {J.-B.}\ \bibnamefont
  {Zuber}},\ }\href@noop {} {\emph {\bibinfo {title} {Quantum field theory}}}\
  (\bibinfo  {publisher} {Courier Corporation},\ \bibinfo {year}
  {2012})\BibitemShut {NoStop}%
\bibitem [{\citenamefont {Di~Piazza}(2008)}]{di2008exact}%
  \BibitemOpen
  \bibfield  {author} {\bibinfo {author} {\bibfnamefont {A.}~\bibnamefont
  {Di~Piazza}},\ }\bibfield  {title} {\bibinfo {title} {Exact solution of the
  landau-lifshitz equation in a plane wave},\ }\href@noop {} {\bibfield
  {journal} {\bibinfo  {journal} {Letters in Mathematical Physics}\ }\textbf
  {\bibinfo {volume} {83}},\ \bibinfo {pages} {305} (\bibinfo {year}
  {2008})}\BibitemShut {NoStop}%
\bibitem [{\citenamefont {Hadad}\ \emph {et~al.}(2010)\citenamefont {Hadad},
  \citenamefont {Labun}, \citenamefont {Rafelski}, \citenamefont {Elkina},
  \citenamefont {Klier},\ and\ \citenamefont {Ruhl}}]{hadad2010effects}%
  \BibitemOpen
  \bibfield  {author} {\bibinfo {author} {\bibfnamefont {Y.}~\bibnamefont
  {Hadad}}, \bibinfo {author} {\bibfnamefont {L.}~\bibnamefont {Labun}},
  \bibinfo {author} {\bibfnamefont {J.}~\bibnamefont {Rafelski}}, \bibinfo
  {author} {\bibfnamefont {N.}~\bibnamefont {Elkina}}, \bibinfo {author}
  {\bibfnamefont {C.}~\bibnamefont {Klier}},\ and\ \bibinfo {author}
  {\bibfnamefont {H.}~\bibnamefont {Ruhl}},\ }\bibfield  {title} {\bibinfo
  {title} {Effects of radiation reaction in relativistic laser acceleration},\
  }\href@noop {} {\bibfield  {journal} {\bibinfo  {journal} {Physical Review
  D}\ }\textbf {\bibinfo {volume} {82}},\ \bibinfo {pages} {096012} (\bibinfo
  {year} {2010})}\BibitemShut {NoStop}%
\bibitem [{\citenamefont {Elkina}\ \emph {et~al.}(2014)\citenamefont {Elkina},
  \citenamefont {Fedotov}, \citenamefont {Herzing},\ and\ \citenamefont
  {Ruhl}}]{elkina2014improving}%
  \BibitemOpen
  \bibfield  {author} {\bibinfo {author} {\bibfnamefont {N.}~\bibnamefont
  {Elkina}}, \bibinfo {author} {\bibfnamefont {A.}~\bibnamefont {Fedotov}},
  \bibinfo {author} {\bibfnamefont {C.}~\bibnamefont {Herzing}},\ and\ \bibinfo
  {author} {\bibfnamefont {H.}~\bibnamefont {Ruhl}},\ }\bibfield  {title}
  {\bibinfo {title} {Improving the accuracy of simulation of radiation-reaction
  effects with implicit runge-kutta-nystr{\"o}m methods},\ }\href@noop {}
  {\bibfield  {journal} {\bibinfo  {journal} {Physical Review E}\ }\textbf
  {\bibinfo {volume} {89}},\ \bibinfo {pages} {053315} (\bibinfo {year}
  {2014})}\BibitemShut {NoStop}%
\end{thebibliography}%

\end{document}